%% file: nima_isodar_2018.tex
\documentclass[5p]{elsarticle}

\usepackage{lineno,hyperref}
\usepackage{booktabs}
\usepackage{wasysym}
\modulolinenumbers[250]

\journal{Nuclear Instruments and Methods in Physics Research Section A}

\usepackage{siunitx}
\usepackage{xspace}
\usepackage{xcolor}
\newcommand{\degree}{\ensuremath{^\circ}\xspace}
\newcommand{\htp}{\ensuremath{\mathrm{H}_2^+}\xspace}
\newcommand{\hthp}{\ensuremath{\mathrm{H}_3^+}\xspace}

\newcommand{\BE}[0]{\begin{equation}}
\newcommand{\EE}[0]{\end{equation}}
\newcommand{\BEA}[0]{\begin{eqnarray}}
\newcommand{\EEA}[0]{\end{eqnarray}}

\mathchardef\mhyphen="2D

\newcommand{\figref}[1]{Figure~\ref{#1}}
\newcommand{\tabref}[1]{Table~\ref{#1}}
\newcommand{\secref}[1]{Section~\ref{#1}}

\newcommand{\DD}{DAE$\delta$ALUS\xspace}

\begin{document}
\begin{frontmatter}

\title{High intensity cyclotrons for neutrino physics}

\author[mit_address]{Daniel Winklehner\corref{mycorrespondingauthor}}
\cortext[mycorrespondingauthor]{Corresponding author}
\ead{winklehn@mit.edu}

\author[mit_address]{Jungbae Bahng}
\author[infnaddress]{Luciano Calabretta}
\author[infnaddress]{Alessandra Calanna}
\author[veccaddress]{Alok Chakrabarti}
\author[mit_address]{Janet Conrad}
\author[infnaddress]{Grazia D'Agostino}
\author[veccaddress]{Siddharta Dechoudhury}
\author[veccaddress]{Vaishali Naik}
\author[mit_address]{Loyd Waites}
\author[drexaddress]{Philip Weigel}

\address[mit_address]{Massachusetts Institute of Technology, 77 Massachusetts Ave, 
                     Cambridge, MA 02139, USA}
\address[drexaddress]{Drexel University, 3141 Chestnut Street, Philadelphia, 
                        PA 19104, USA}
\address[infnaddress]{INFN, Via S. Sofia 62, 95123 Catania, Italy}              
\address[veccaddress]{VECC \& HBNI, Sector-1, Block-AF, Bidhan Nagar, Kolkata 700064, India}  

%
%

\begin{abstract}
In recent years, the interest in high intensity proton beams in excess of 
several milli-amperes has risen. Potential applications are in neutrino physics, 
materials and energy research, and isotope production. Continuous wave proton 
beams of five to ten milli-amperes are now in reach due to advances in 
accelerator technology and through improved understanding of the beam dynamics.
As an example application, we present the proposed IsoDAR experiment, a search for
so-called sterile neutrinos and non-standard interaction using the KamLAND detector 
located in Japan. We present updated sensitivities for this experiment and 
describe in detail the design of the high intensity proton driver that uses 
several novel ideas. These are: accelerating \htp instead of protons, 
directly injecting beam into the cyclotron via a radio frequency quadrupole, 
and carefully matching the beam to achieve so-called vortex motion.
The preliminary design holds up well in PIC simulation studies and the injector
system is now being constructed, to be commissioned with a \SI{1}{MeV/amu} test cyclotron.
\end{abstract}


\end{frontmatter}

\linenumbers

\input{Sec1_Introduction}

\input{Sec2_AcceleratorDesign}

\input{Sec3_Simulations}

\input{Sec4_Conclusion}

\section{Acknowledgements}
MIST-1 is supported by NSF grant \#PHY-1505858 and funding from the Bose Foundation. The RFQ-DIP project is supported by NSF grant \#PHY-1626069 and the
Heising-Simons Foundation.
The authors are very thankful for the support of the MIT Central Machine shop, the 
MIT Plasma Science and Fusion Center (PSFC), and the University of Huddersfield 
for support with machining, lab space and utilities, and equipment, respectively.
The corresponding author would also like to thank the American Physical Society
and the Indo-U.S. Science and Technology Forum for awarding him a travel grant 
through the Ph.D. student and Postdoc Visitation Program to
go to Kolkata and collaborate on RFQ design with the RIB group at VECC.

\section*{References}

\bibliography{nima_isodar_2018}

\end{document}

%% file: Sec1_Introduction.tex
\section{Introduction}
High intensity proton beams are used very successfully in spallation
neutron sources like the Spallation Neutron Source (SNS) at Oak Ridge National Laboratory (ORNL) \cite{mason:sns}, the Materials and Life Science Experimental Facility (MLF) at the Japan Proton Accelerator Research Complex (J-PARC) 
\cite{maekawa:jsns}, and the Swiss spallation neutron source SINQ at the 
Paul Scherrer Institute (PSI) \cite{fischer:sinq}. 
In addition to neutrons these spallation targets can also produce kaons, pions, muons, 
and neutrinos to be used in particle physics experiments (e.g. \cite{akimov:coherent}).
With the exception of PSI, these facilities use linear accelerators to produce high intensity proton beams,
which are costly and require significant space. 
Recent developments in accelerator technology and improved understanding of the 
beam dynamics in high intensity cyclotrons have made these circular particle accelerators an attractive candidate for new high intensity proton drivers.
Currently, PSI holds the record of highest power continuous wave (cw) proton beam from 
a cyclotron with \SI{2.2}{mA} at \SI{590}{MeV} energy, which amounts to $\approx 1.3$ MW of beam power \cite{seidel:psi_status}. In this coupled cyclotron system, so-called \emph{vortex motion} has been observed experimentally and been reproduced in Particle-In-Cell (PIC) simulations
\cite{stammbach:vortex, yang:cyclotron_sim}. Here, the external focusing forces 
of the isochronous cyclotron combined with space-charge cause a spiraling of the 
bunch that can lead to longitudinal focusing. We can utilize this effect to accelerate even higher beam currents as will be discussed in this paper. Some proposed experiments and potential applications of high intensity proton beams are listed in \tabref{tab:uses} together with current and energy requirements.

\begin{table}[!h]
	\vspace{-10pt}
	\footnotesize
	\caption{A few potential uses for high current proton beams.
    (ADSR - Accelerator Driven Sub-critical Reactors,
    ADS - Accelerator Driven Systems)}
	\label{tab:uses}
	\centering
    \vspace{5pt}
    \renewcommand{\arraystretch}{1.25}
		\begin{tabular}{llll}
            \hline
            \textbf{Experiments} & \textbf{Type} & \textbf{Current}  & \textbf{Energy}\\
            \hline \hline
            IsoDAR \cite{IsoDARsterilePRL, abs:isodar_cdr1} & neutrino exp. 
            & 10~mA & 60~MeV \\
            \DD \cite{abs:daedalus, aberle:daedalus} & neutrino exp. 
            & 10~mA & 800~MeV \\
            ADSR \cite{ishi:adsr,rubbia:adsr} & energy & 10~mA & $<1.2$~GeV \\
            ADS \cite{biarrotte:ads,lisowski:ads} & energy & 4-120~mA & $<1.2$~GeV \\
            Isotopes \cite{schmor:isotopes,alonso:isotope} & medicine & $<2$~mA 
            & 3-70~MeV \\
            \hline
		\end{tabular}
\end{table}

After motivating the need for high intensity proton beams by using one of the 
given examples (the proposed neutrino experiment IsoDAR) in the following subsection, 
we will describe in detail the building blocks of a cyclotron-based proton driver 
in \secref{sec:accel_design} together with the design choices that make such a 
system feasible. In \secref{sec:simulations} we will then present simulations 
of the individual parts, followed by a discussion of the next steps towards a 
running system in \secref{sec:outlook}.

\subsection{An Example: The IsoDAR Experiment \label{DAR}}

IsoDAR (Isotope Decay At Rest) is a
novel, pure $\bar \nu_e$ source under development that makes use of a cyclotron-accelerated beam delivered to a decay-at-rest target.
A high-intensity H$_2^+$ ion source feeds  a 
60 MeV/amu cyclotron. The beam is electron-stripped after extraction and
transported to a
$^9$Be target, producing neutrons.  The neutrons enter 
a $\ge$99.99\% isotopically pure $^7$Li sleeve, where neutron
capture results in $^8$Li.  The  $^8$Li undergoes
$\beta$ decay-at-rest, producing an isotropic, pure
$\bar \nu_e$ flux.
Pairing this very high-intensity $\bar \nu_e$ source with a detector that contains hydrogen 
allows for the inverse beta decay (IBD) interaction, $\bar \nu_e
+ p \rightarrow e^+ + n$, and  $\bar \nu_e$-$e^-$ elastic scattering (ES)
processes.  

Our goal in developing IsoDAR was to ``think outside of the box'' on
neutrino sources for underground science.   We are producing a 
cheaper, more
compact and purer source than existing neutrino sources, with
flexibility to install the source underground at most laboratories.

Our initial proposal is to install IsoDAR near to the KamLAND
detector.    The first proposal for the layout of the IsoDAR facility
placed the source at 16 m from the center of the KamLAND detector and
assumed an energy resolution of 6.5\%/$\sqrt{E}$
\cite{abs:isodar_cdr1}.  Recently, two potential improvements to this plan have
been put forward.    First, we have 
identified a new, potentially closer targeting site \cite{alonso:isodar_cdr2}, at
12.5 m, increasing rates by $\times 1.6$.   
Should this new location be employed, the additional statistics can allow us to
either reach our physics goals more quickly or to run the cyclotron at 6
mA of protons (3 mA H$_2^+$) rather than 10 mA of protons (5 mA
H$_2^+$) for the original time period.   Running with 6 mA of protons is preferred to reduce
operations cost and risk.  Second, KamLAND is proposing an upgrade
which will improve the resolution of the detector to 
3\%/$\sqrt{E}$,  which benefits IsoDAR physics.   

Because IsoDAR is relatively inexpensive, one can envision many such
sources in the future.   For example, the JUNO collaboration has worked
with us on potentially bringing IsoDAR to their detector after reactor
running \cite{MIT13}.

\begin{figure}[t]
\begin{center}
\includegraphics[width=.4\textwidth]{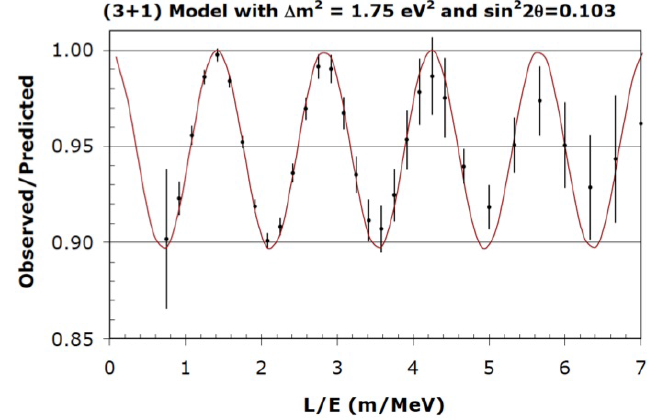} \\
\vspace{5pt}
\includegraphics[width=.4\textwidth]{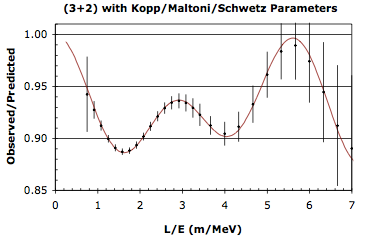}
\end{center}
\caption{\footnotesize IsoDAR$@$KamLAND $L/E$ dependence, 5 years of running,
  for one (upper plot) and two (lower plot) sterile neutrinos. Solid curve is the
  oscillation probability with no smearing in the reconstructed position
  and energy and the data points with error bars are from simulated events
  including smearing from reconstruction for the original target
  location and detector (assumes present resolution, and
  target at \SI{16}{m}).
\label{osc3N}}
\end{figure}

\subsection{Examples of the Physics of IsoDAR}
In this section, we describe two examples of the IsoDAR physics
program in detail.    IsoDAR also makes other beyond Standard Model
searches, nuclear physics measurements
and provides calibration for KamLAND.   

\subsubsection{Sterile Neutrino Search}
Interest in light sterile neutrinos has arisen from
anomalies observed in a 
wide range of short-baseline (SBL) experiments employing neutrinos and antineutrinos of
different flavors and different energies \cite{LSND, MBnu, MBnubar,
  reactor1, reactor2, SAGE3, GALLEX3}.     However,
other experiments
 potentially sensitive to sterile neutrinos have observed null results  \cite{KARMEN, ConradShaevitz, MBNumi,
  MB_SB, MBnudisapp, NOMAD1, CCFR84, CDHS}.  This 
limits models that describe the anomalies.    One such model that introduces one additional sterile
neutrino (``3+1'') depends on 
four BSM parameters, $\sin^2 2\theta_{ee}$,
 $\sin^2 2\theta_{\mu\mu}$,  $\sin^2 2\theta_{\mu e}$, and $\Delta
 m^2$. The first three are the mixing angles measured in three
 types of oscillation experiments:
 $\nu_e$ disappearance, $\nu_\mu$ disappearance and
$\nu_\mu \rightarrow \nu_e$ appearance.  These are
constructed of two matrix elements, $|U_{e4}|^2$ and $|U_{\mu 4}|^2$,
hence they are interdependent.  The fourth parameter, the squared mass
 splitting between the mostly sterile and mostly active states must be
 consistent for all three types of oscillations.

Global fits that include all of the SBL data sets show a
marked improvement in $\Delta \chi^2_{null-min}/\Delta$dof of 50.61/4 
for a 3+1 model \cite{MIT19}.     However, 
the appearance-only and disappearance-only allowed regions show poor
overlap in allowed parameters \cite{MIT21}.  This well-known tension is one of the
primary reasons scientists question the sterile neutrino
model.   

IsoDAR was developed to allow a sterile
neutrino search that can resolve the confusing situation of anomalies
in the electron-flavor sector.    IsoDAR provides a very well-understood flux and a highly
sensitive method to search for sterile neutrinos through
reconstruction of the $L/E$ dependence of the neutrino oscillation process,
commonly referred to as an ``oscillation wave,'' shown in \figref{osc3N}.   

\begin{figure}[t!]
\includegraphics[width=0.9\columnwidth]{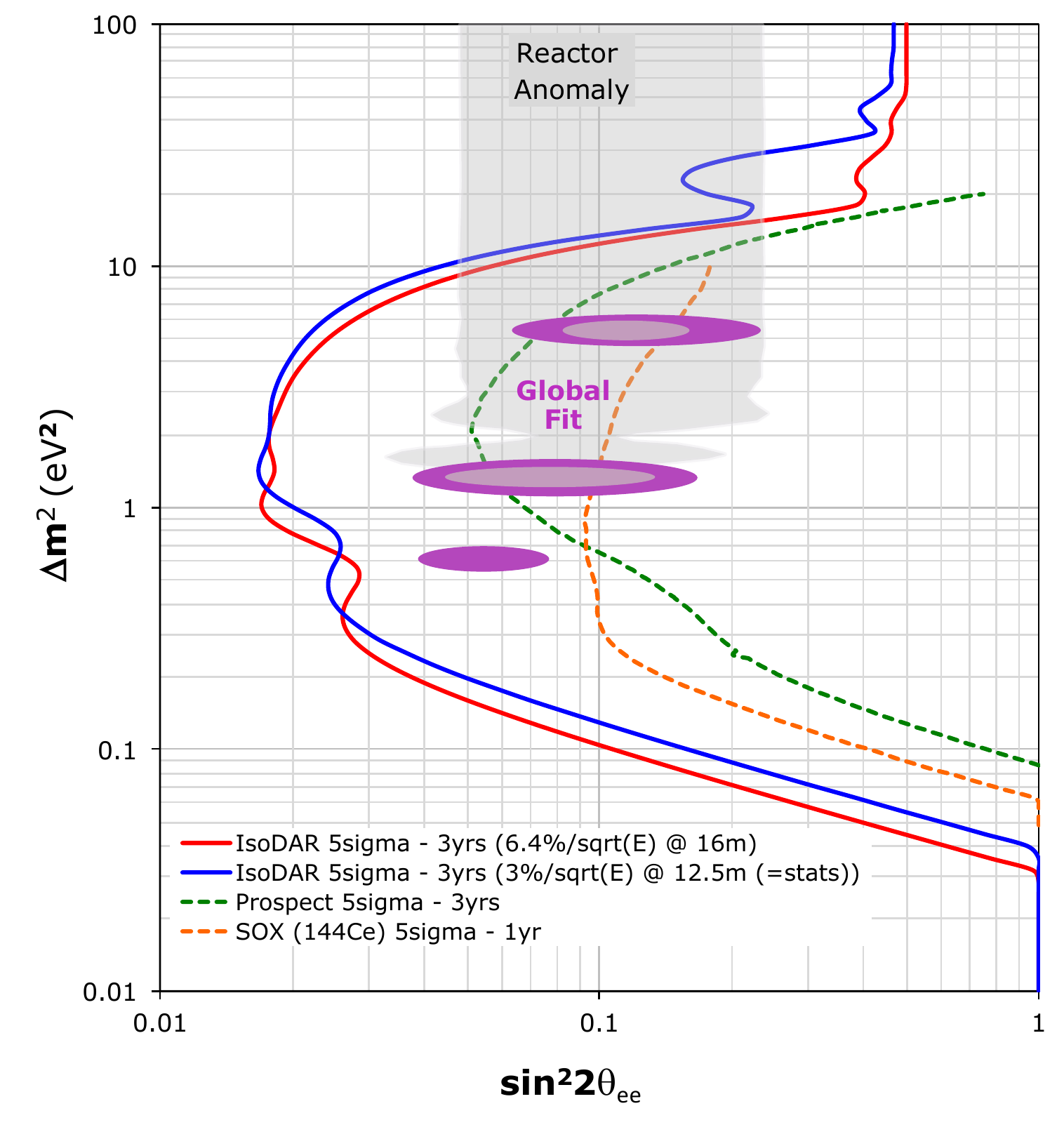}
\caption{\footnotesize \label{isonew} Isodar$@$KamLAND at original
site (red), new proposed site with equal statistics (blue), and in
comparison to other experiments.}
\end{figure}

The IsoDAR 3-year, 5$\sigma$ expectations for our baseline design (red)
\cite{abs:isodar_cdr1} and for equal statistics at the new, closer site, with the
detector upgrade (blue) \cite{alonso:isodar_cdr2} are shown in \figref{isonew}, 
compared to other proposed experiments (green, orange) \cite{Prospect, wurm:sox1}.
IsoDAR$@$KamLAND addresses the reactor anomaly \cite{reactor1}, and
fully covers of all allowed islands from the 3+1 global fit (magenta)\cite{MIT20}.  
\figref{osc3N} shows the $L/E$ distribution for two potential signals.
The well-measured  oscillation wave differentiates models with one sterile neutrino (left)
and two sterile neutrinos (right) \cite{IsoDARsterilePRL}. 

The well-understood $^8$Li flux is a great advantage for IsoDAR
compared with the ongoing reactor experiments.
Reactor fluxes have two important issues that the IsoDAR flux avoids.  The first is an excess in the reactor spectrum at 5 MeV
that is seen in most \cite{DC5MeV, RENO5MeV, DB5MeV}, but not all \cite{Bugey}
cases, for which the source is unknown.    Second, although Daya Bay has
indicated the reactor anomaly may depend on fuel (U vs. Pu) \cite{DBflux}, 
three independent analyses \cite{Schwetzreact, Hayesreact,
Giuntireact} reach the conclusion that even after the new Daya Bay
result, the reactor anomaly
remains at $>2\sigma$.
IsoDAR does not face these questions.

\subsubsection{NSI Search Through Precision Electroweak Tests}
The search for indications of NSIs is a second example of 
the outstanding new physics opened by the IsoDAR source. 
This makes use of the fact that, running with KamLAND, the experiment will collect the largest sample of $\bar\nu_e$ ES events observed to date
\cite{nue-elastic}.  
Approximately 2600 ES events will be collected at KamLAND above a 3~MeV visible
energy threshold over a 5 year run--a factor of five above existing
$\bar \nu_e$
samples \cite{nue-elastic}.  The improved
resolution of KamLAND may allow us to lower this threshold and
increase statistics further.  
The ES rate can be normalized
using the well-understood, very-high-statistics IBD events.       
Ref.~\cite{nue-elastic} describes how 
the ES cross section can be used to measure the weak couplings,
$g_V^{\nu_e e}$
and $g_A^{\nu_e e}$.  Deviation of these couplings could explain the
NuTeV anomaly \cite{NuTeV}.   These couplings can be recast  in terms of the left
and right handed couplings that may be modified by NSI's:  $g_L^{SM} \rightarrow g_L^{SM} +
 \varepsilon^{eL}_{ee}$ and/or $g_R^{SM} \rightarrow g_R^{SM} +
 \varepsilon^{eR}_{ee}$. 
IsoDAR's capability to constrain NSI's  is as strong as the combined
present data sets, and is orthogonal because this is a pure antineutrino sample.

%% file: Sec2_AcceleratorDesign.tex
\section{Accelerator Design\label{sec:accel_design}}
\begin{figure}[t!]
	\centering
		\includegraphics[width=1.0\columnwidth]
        {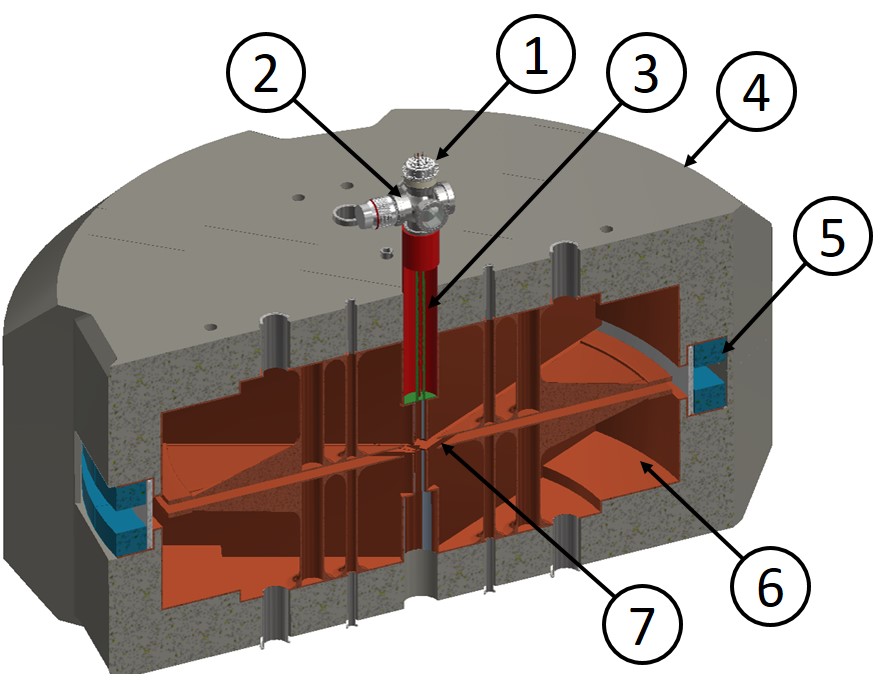}
	    \caption{Conceptual layout of the compact cyclotron driver for high intensity
                 60 MeV/amu \htp beams.
                 1. Ion source,
                 2. diagnostic and pumping box,
                 3. Radio Frequency Quadrupole (RFQ) injector,
                 4. Compact cyclotron yoke,
                 5. Compact cyclotron coils,
                 6. Accelerating RF electrodes (``Dees''),
                 7. Central region.} 
	    \label{fig:accel_layout}
\end{figure}
As mentioned in the introduction, in order for IsoDAR@KamLAND to be decisive in 
5 years, a current of 10 mA of protons on target (6 mA in the alternate location) is necessary. This is roughly a factor 4 more than world-record holder PSI Injector II has
delivered so far. The main limitation being space-charge 
(the de-focusing self-fields of the beam resulting from Coulomb-repulsion 
between the individual particles).
Space charge effects are most pronounced at low energies and high current densities, 
i.e. in the Low Energy Beam Transport line (LEBT) connecting ion source and 
accelerator, the injection into the accelerator, and during the first few turns in
the cyclotron, until higher beam energy has been reached.
In order to mitigate these detrimental space charge effects, we propose to 
accelerate \htp ions instead of protons and, at the same time,
increase the injection energy of the beam to \SI{35}{keV/amu}. 
Using \htp allows the use of a stripper foil after extraction from the cyclotron 
to remove the binding electron, thereby doubling the electrical beam current.
This reduces our requirement to 5 mA of \htp (3 mA in the alternate location).
As will be explained in \secref{sec:spiral}, using \htp leads to a fairly large 
spiral inflector (the electrostatic device bending the beam from the axial
direction onto the cyclotron median plane).

Another limiting factor is the phase acceptance of the cyclotron. Typically, for 
isochronous cyclotrons this is $\pm 10\degree$ around the synchronous phase $\Phi_\mathrm{S}$. Our preliminary beam dynamics studies for the \SI{60}{MeV/amu}
cyclotron have shown that the vortex effect can lead to a matched beam in later
stages of acceleration even if the injected beam is as long as $\pm 20\degree$,
if carefully placed collimators are used to clean up beam halo in the first 
10 turns (see \cite{yang:daedalus, jonnerby:thesis}). 
Even so, a significant portion of the injected beam will 
be lost. Our solution is to use a Radio Frequency Quadrupole (RFQ) to aggressively 
pre-bunch the beam.
In order to bring the RFQ as close as possible to the central region of the cyclotron
(injection point), it will be embedded partially in the iron yoke of the cyclotron.
The conceptual layout of the proton driver is shown in \figref{fig:accel_layout}.
The main components of the driver are:
\begin{enumerate}
\item The ion source
\item The RFQ injector
\item The spiral inflector
\item The cyclotron
\end{enumerate}
In order to demonstrate the \htp production and RFQ bunching/injection capabilities,
an NSF funded project is currently on the way at MIT, the \emph{RFQ-Direct Injection 
Project} (RFQ-DIP), which combines items 1-3 with a \SI{1}{MeV/amu} test cyclotron.
RFQ-DIP will also double as the actual ion source and RFQ for IsoDAR.

In the following subsections, we describe the individual components of the proton
driver in more detail. For the full cyclotron design, we will mostly refer to 
previous publications (references given in \secref{sec:cyclo}) and will focus on 
the central region and the RFQ-DIP \SI{1}{MeV/amu} test cyclotron design. 
However, these considerations are directly applicable in the full 
\SI{60}{MeV/amu} machine.

\input{Sec2.1_MIST}

\input{Sec2.2_RFQ}

\input{Sec2.3_Cyclotron}

\input{Sec2.4_Spiral}

%% file: Sec2.1_MIST.tex
\subsection{Ion source design\label{sec:source}}
In order to produce the required DC current of \htp (10-15 mA), a new
multicusp ion source (MIST-1) has been constructed at MIT and is currently 
under commissioning. 
The new source is based on a design by Lawrence Berkeley National Laboratory 
dedicated to the production of \htp over protons which demonstrated combined 
current densities of $50~\mathrm{mA}/\mathrm{cm}^2$ and an \htp fraction greater 
than 80\% \cite{ehlers:multicusp1}.

\subsubsection{Ion source design parameters}
\begin{figure}[t!]
	\centering
		\includegraphics[width=0.9\columnwidth]
        {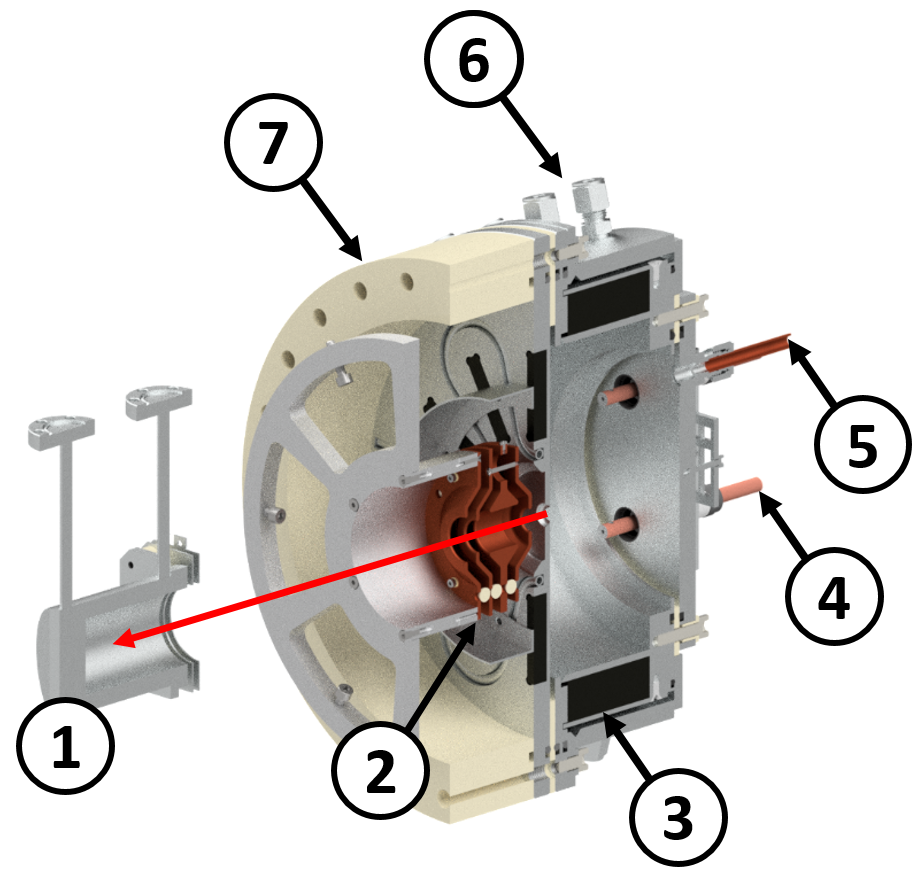}
	    \caption{Cut view of the MIST-1 ion source.
	             1. Faraday cup, 2. Extraction System,
	             3. Permanent magnets (Sm$_2$Co$_{17}$),
	             4. Filament feedthroughs,
	             5. Gas inlet,
	             6. Water cooling fittings,
	             7. Alumina insulator ring.}
	\label{fig:source}
\end{figure}

A detailed description of the MIST-1 ion source can be found 
elsewhere \cite{abs:isodar_cdr1, axani:mist1}. In short, a plasma 
is confined by permanent magnets (Sm$_2$Co$_{17}$), electrons are produced
by a filament and are accelerated by an externally 
applied electric field of order \SI{100}{V}. Due to the short source and the 
plasma being produced close to the extraction region, there is not enough time for
dissociation to occur, favoring \htp over proton production. A half-section
view of the ion source can be seen in \figref{fig:source} and the main parameters 
are listed in \tabref{tab:source}.

\begin{table}[!b]
    \centering
    \caption{MIST-1 ion source parameters. \label{tab:source}}
    \begin{tabular}{ll}
        \toprule
        \textbf{Parameter} & \textbf{Value (nominal)}\\
		\midrule
            Plasma chamber length & 6.5 cm\\
            Plasma chamber diameter & 15 cm \\
            Permanent magnet material & Sm$_2$Co$_{17}$\\
            Permanent magnet strength & 1.05 T on surface\\
            Front plate magnets & 12 bars (star shape)\\
            Radial magnets & 12 bars \\
            Back plate magnets & 4 bars, 3 parallel rows\\
            Front plate cooling & embedded steel tube  \\ 
            Back plate cooling & embedded copper tube\\
            Chamber cooling & water jacket\\
            Water flow (both) & (1.5 l/min)\\
            Filament feedthrough cooling & air cooled heat sink \\
            Filament material & 98\% W, 2\% Th\\
            Filament diameter & $\approx 1.5$ mm  \\
            Discharge voltage & max. 150 V\\
            Discharge current & max. 24 A \\
            Filament heating voltage & max. 8 V\\
            Filament heating current & max. 100 A\\
        \bottomrule
    \end{tabular}
\end{table}

\subsubsection{Preliminary commissioning results}
In the first commissioning phase, a thinner (\SI{0.4}{mm} diameter) pure tungsten 
filament was used instead of the nominal filament described in the previous section. Currents were measured in a Faraday cup right after the extraction system (see \figref{fig:source}), thus not allowing species separation. All reported currents are total extracted currents.
During the accumulated run time of $\approx 30$ hours, 
the source showed good stability for about 4 hours at a time, reaching a maximum current density of 16 mA/cm$^2$ (4.6 mA total) \cite{winklehner:mist1}. While 
at this point the commissioning results are preliminary, we are confident that
our source will have similar species ratios 
($\mathrm{p}^+:\htp:\textrm{H}_3^+\approx 1:8:1$)
as previous similar sources \cite{ehlers:multicusp1}. A paper on the final
commissioning results using a short analysis beam line with dipole magnet
for mass separation is forthcoming.

\subsubsection{Extraction system}

\begin{figure}[t!]
	\centering
		\includegraphics[width=1.0\columnwidth]
        {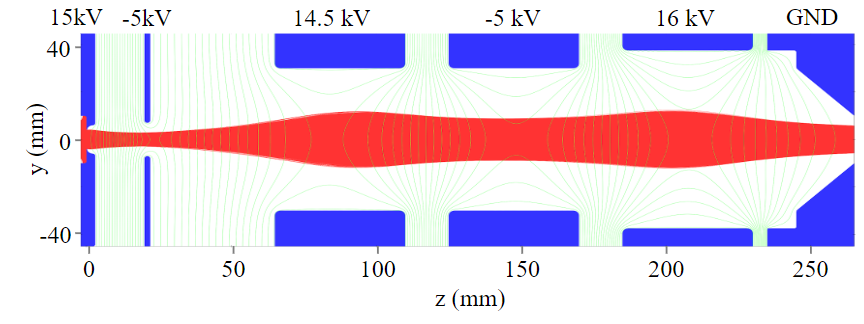}
	    \caption{ 2D cut view from center of extraction system of ion beam (red) moving through the extraction system before being transferred into the RFQ.}
	    \label{fig:extraction}
\end{figure}

\begin{figure}[t!]
	\centering
		\includegraphics[width=0.8\columnwidth]
        {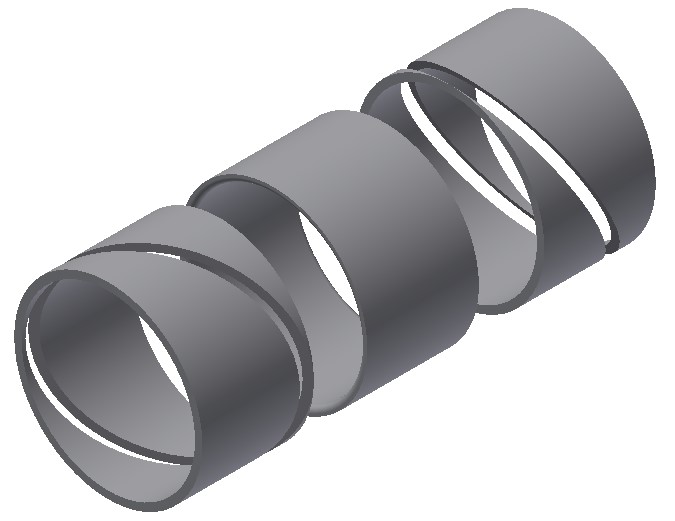}
	    \caption{Cartoon picture of the proposed segmentation of lens 
                 elements 1 and 3 to allow small angle steering for additional
                 control over matching the beam into the RFQ.}
	    \label{fig:slits}
\end{figure}

The extraction system serves two important purposes: initial shaping of the beam,
and focusing the beam into the RFQ. The extraction system designed for the ion
source commissioning (item 2 in \figref{fig:source}) is very short and not 
able to serve in the second function. For the final IsoDAR system and the 
intermediate RFQ-DIP, a new extraction system is being designed.
A cross-sectional view is shown in \figref{fig:extraction}. 
The electrodes (from left to right) are: plasma electrode (+15 kV), puller electrode (-3 to -15 kV), lens element 1 (+10 to +15 kV), lens element 2 (-3 to -15 kV), lens element 3 (+10 to +15 kV), and RFQ entrance flange (ground). 
The extraction system was modeled using the 
well-established IBSIMU code \cite{kalvas:ibsimu}.
IBSIMU calculates the ion trajectories in an iterative process, taking into 
account the (quasi-neutral) ion source plasma,
the external fields generated by the electrodes, and the beam's self-fields
(space charge).
In the final design, lens elements 1 and 3 will be segmented (given diagonal slits similar to Ref. \cite{Mandal:einzel}) to allow small angle steering in both 
horizontal and vertical direction. 
A simplified picture of this scheme is shown in \figref{fig:slits}.

%% file: Sec2.2_RFQ.tex
\subsection{RFQ Design}
In order to improve injection efficiency into a compact isochronous cyclotron, 
the maximum possible pre-bunching is desired, as the accelerator can only accept 
beam into a phase window of $20-40\degree$
An RFQ provides over 90\% bunching efficiency, up to 100\% transmission 
efficiency, and acts as a pre-accelerator. Unwanted ion species like protons and $\mathrm{H}_3^+$ are lost to the electrodes and the cavity wall.
In the following subsections the preliminary design of the IsoDAR/RFQ-DIP 
RFQ is discussed. A first start-to-end simulation from ion source extraction 
through the RFQ and spiral inflector, 
up to acceleration to 1 MeV/amu in the cyclotron is presented in 
\secref{sec:simulations}.

\subsubsection{RFQ design considerations}
\begin{table}[!b]
    \centering
    \caption{Requirements for the RFQ design. \label{tab:rfq_reqs}}
    \begin{tabular}{lrl}
        \toprule
        \textbf{Parameter} & \textbf{Value (nominal)} & \textbf{Unit}\\
		\midrule
            Frequency &	32.8 & MHz \\
		    Input energy & 15 & keV \\
            Length & $<1500$ & mm \\
            Particle & \htp & (q/A = 1/2) \\
            Output energy & 70 & keV \\
            Diameter & $<300$ & mm \\
        \bottomrule
    \end{tabular}
\end{table}
The design requirements for the RFQ are listed in \tabref{tab:rfq_reqs}.
These requirements for the IsoDAR/RFQ-DIP RFQ are driven by the following
considerations: A compact cyclotron was chosen as the main accelerator 
due to space restrictions near underground neutrino detectors. 
Thus axial injection is the only viable solution to inject beam into the 
cyclotron. To avoid unnecessary beam spreading between RFQ exit and 
spiral inflector entrance, the RFQ must be brought as close as possible
to the cyclotron median plane.
The frequency must be matched to the cyclotron operation 
frequency (\SI{32.8}{MHz}), the size
(length, diameter) follows directly from the necessity to embed the RFQ 
partially in the cyclotron's iron yoke. The input energy (\SI{7.5}{keV/amu})
is minimized to reduce length (cell length $= v_z/c\cdot\lambda$) and the output 
energy (\SI{35}{keV/amu}) is the 
maximum energy that can safely be injected through a large spiral inflector.
As will be discussed in more detail later, in order to achieve the comparably low operation frequency of \SI{32.8}{MHz} while 
maintaining the required small diameter, a split-coaxial design has been employed.
In order to reduce energy spread and longitudinal beam size at the matching point
(first accelerating gap in the cyclotron central region) a re-bunching cell is 
added to the RFQ.

We followed the usual cycle of beginning with a first order 
physics/beam dynamics design using the simulation software PARMTEQM developed
at Los Alamos National Laboratory (LANL) \cite{crandall:parmteq1}, followed by 
a 3D Particle-In-Cell (PIC) simulation using a fieldmap calculated from 
the PARMTEQM vane/rod shapes. For the PIC calculations we used TRACK 
~\cite{trackcode} and fields were calculated by a Finite Elements Method 
(FEM) in CST Microwave Studio \cite{cstcode} after modeling the 
vanes in Autodesk Inventor.

\subsubsection{RFQ beam dynamics design with PARMTEQM}
\begin{figure}[t!]
	\centering
		\includegraphics[width=1.0\columnwidth]
        {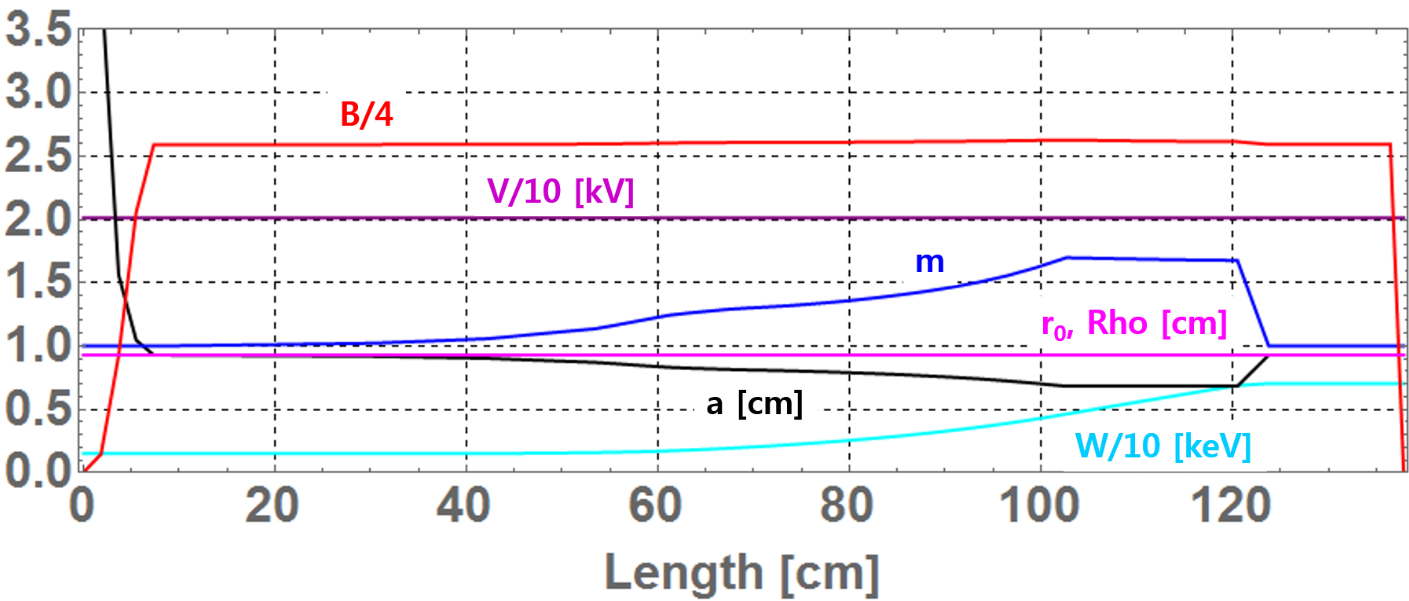}
        \includegraphics[width=1.0\columnwidth]
        {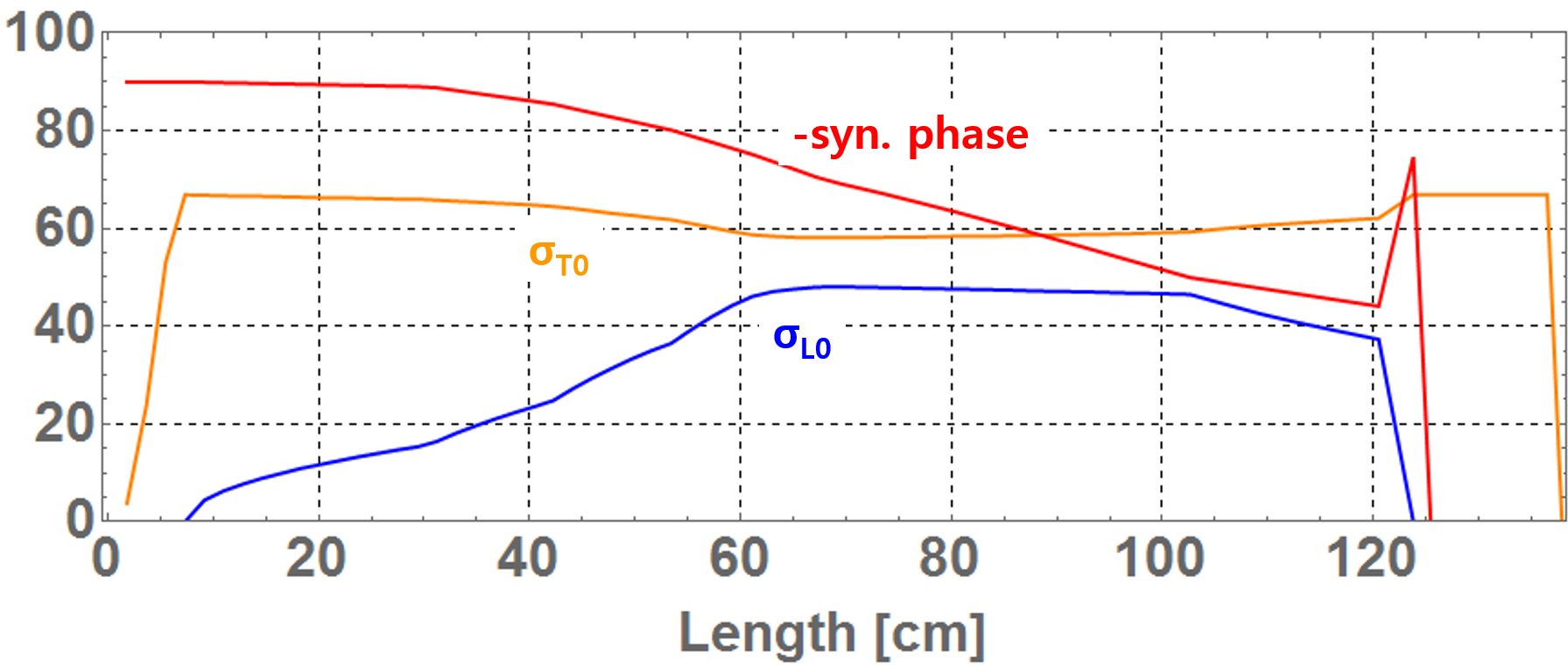}
	    \caption{RFQ Design parameters versus cell number.}
	    \label{fig:rfq_params}
\end{figure}
\begin{table}[!b]
    \centering
    \caption{RFQ design parameters. 
             \label{tab:rfq_params}}
    \begin{tabular}{lrl}
        \toprule
        \textbf{Parameter} & \textbf{Value} & \textbf{Unit}\\
		\midrule
            Frequency &	32.80 & MHz \\
            Length & 1378.69 & mm\\
            Transmission rate & 97.27 & \%\\
            $\epsilon_{\mathrm{transv., n, 4rms}}$ & 1.37 & $\pi$-mm-mrad\\
            a, min. vane-tip & 6.83 & mm\\
            r0, mid-radius & 9.30 & mm\\
            Vane voltage & 20.14 & kV\\
            $\epsilon_{\mathrm{long., 1rms}}$ & 40.24 & keV-deg\\
            $\rho$, vane-tip curvature & 9.30 & mm\\
            Octupole term & 0.07 & \\
            No. of cells & 58 & \\
        \bottomrule
    \end{tabular}
\end{table}
In addition to the fixed requirements of \tabref{tab:rfq_reqs}, it is desirable
have a constant vane voltage along the RFQ for added ease of machining.
The detailed design results of the IsoDAR RFQ are listed in \tabref{tab:rfq_params}
and plotted in \figref{fig:rfq_params}
This preliminary RFQ design features a constant vane voltage of \SI{20.14}{kV}, 
vane-tip curvature $\rho$ of \SI{9.30}{mm} and mid-radius $r_{0}$ of \SI{9.30}{mm}.
This gives a comparably simple structure with easy operation and easy tuning 
as well as a small longitudinal emittance of \SI{40.24}{keV-deg} and 
a high transmission rate of \SI{97.27}{\%}.  
Since modulation and synchronous phase in the low energy region are a key parameter to determine longitudinal emittance, adiabatic varied modulation and synchronous phase were applied as shown in \figref{fig:rfq_params} to obtain a high transmission rate 
and small longitudinal emittance. In addition, we smoothly varied parameters in the shaper and gentle buncher sections to avoid transverse emittance blow-up at the end of 
the bunching section due to strong defocussing term. 

\subsubsection{RFQ geometrical cavity design}
\begin{figure*}[t!]
	\centering
		\includegraphics[height=5.45cm]
        {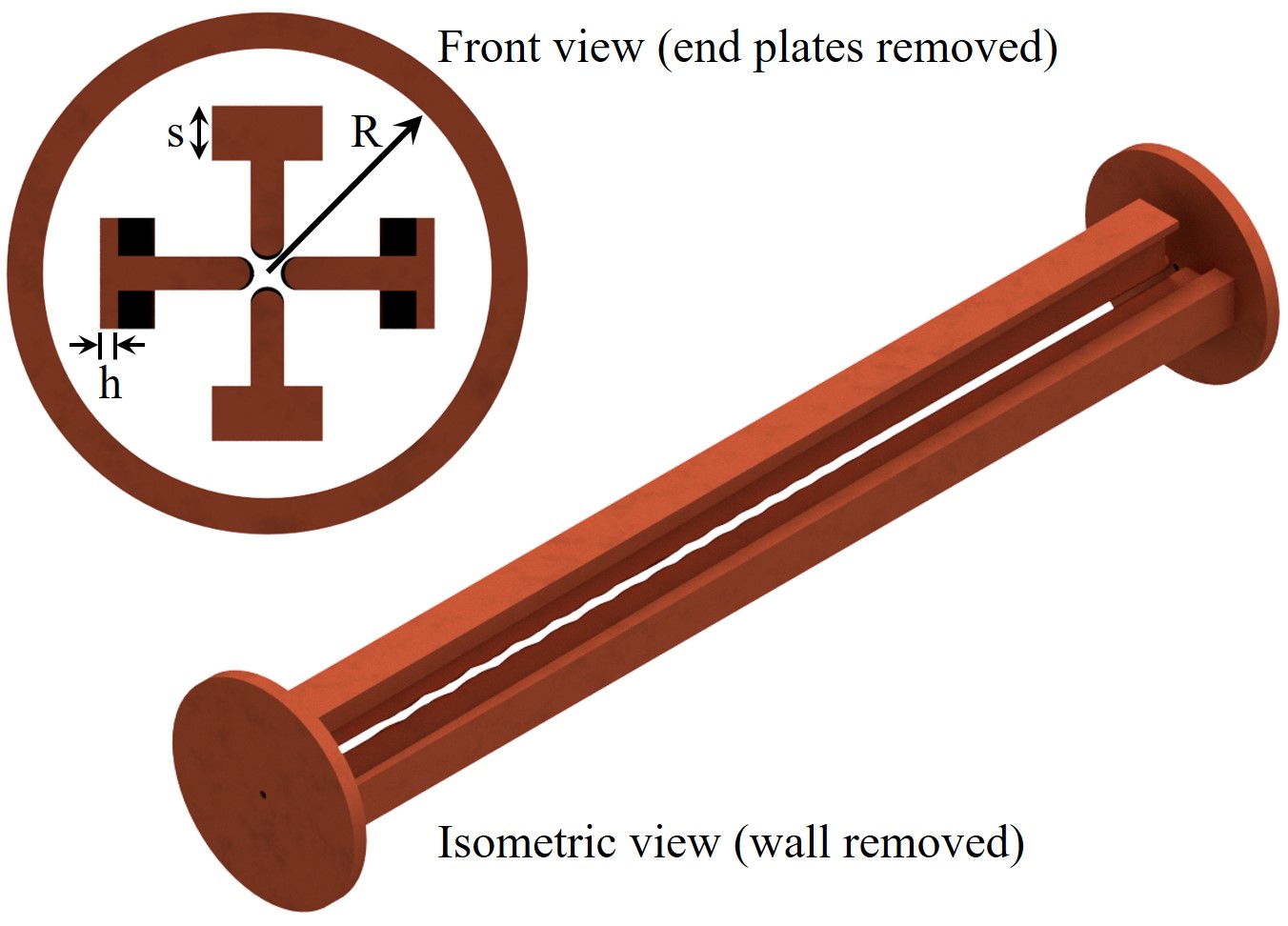}
        \includegraphics[height=5.45cm]
        {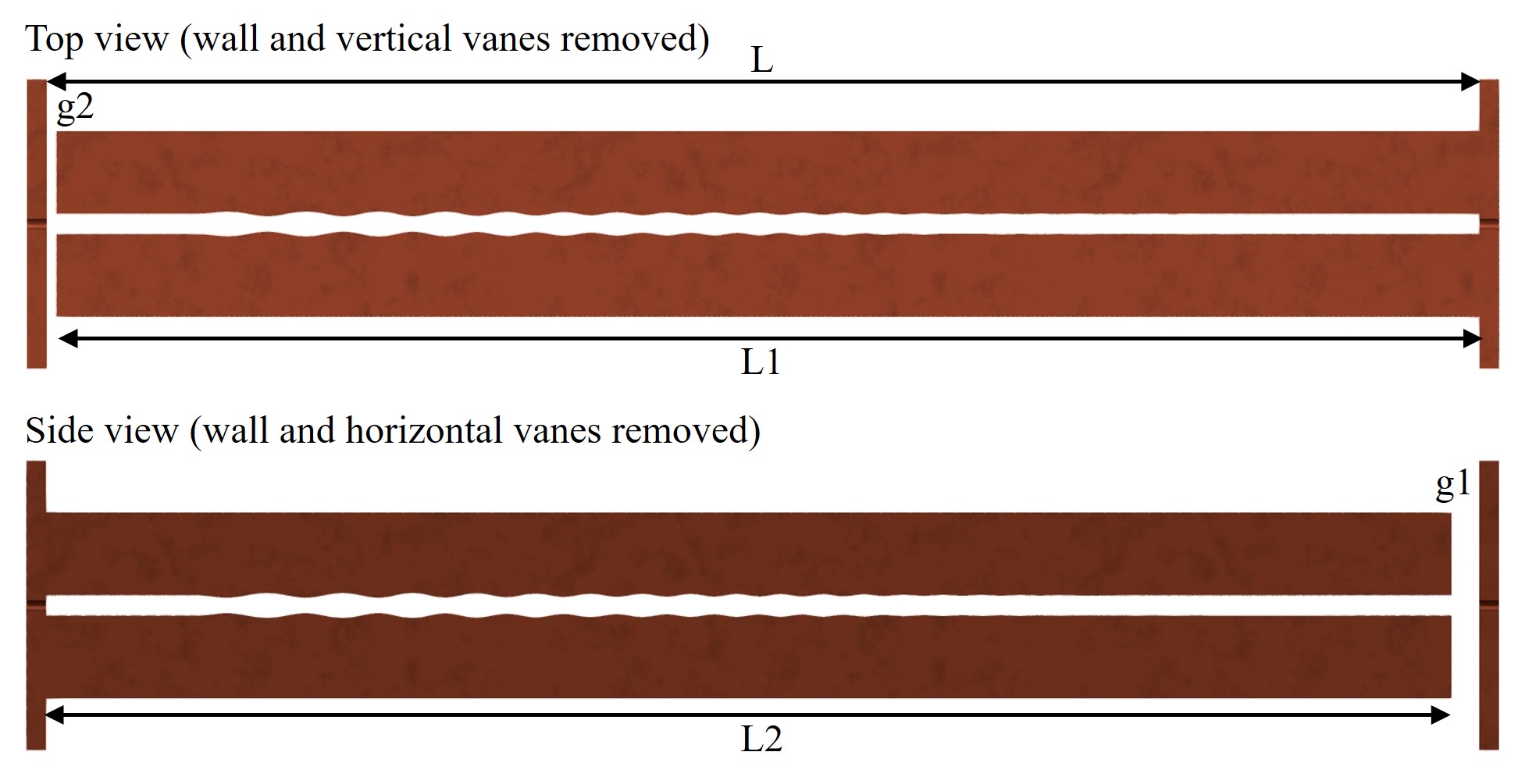}
	    \caption{CAD renderings of the RFQ without re-bunching cell. 
                 The beam enters from the right. The most important geometry
                 parameters are indicated and listed in \tabref{tab:rfq_cav_params}.
                 In the front and isometric views, the tapered skirts are 
                 visible that are used to achieve the design frequency and add
                 additional stability against droop.}
	    \label{fig:rfq_cavity}
\end{figure*}
In general, a four-vane cavity type has a simple shape and high shunt impedance.
Unfortunately, for low frequencies the transverse dimension gets prohibitively
large. As mentioned above, for this project, cavity size is critical.
Instead, a split-coaxial cavity design has been proposed and developed to obtain 
a small diameter at low operation frequency. In the split-coaxial cavity design,
two opposite vanes are connected to one of the end-plates and the two other vanes 
(again opposite vanes) are connected to the other end-plate
(this is depicted in \figref{fig:rfq_cavity}). 
The smaller transverse size in the split-coaxial RFQ cavity is due to the special 
magnetic flux structure \cite{scrfqmerit}. 
In addition to the lower RF frequency, this also provides high stability of the
acceleration field since the magnetic field encompasses all four vanes as a 
coaxial waveguide.
The geometrical cavity parameters for our split-coaxial RFQ cavity are defined as shown in \figref{fig:rfq_cavity} and listed in \tabref{tab:rfq_cav_params}.
The RF frequency for this structure has been calculated to be 32.64 MHz
using CST Microwave Studio.
\begin{table}[!b]
    \centering
    \caption{RFQ cavity geometrical parameters. Select parameters are also shown 
             in \figref{fig:rfq_cavity}. \label{tab:rfq_cav_params}}
    \begin{tabular}{lrl}
        \toprule
        \textbf{Parameter (description)} & 
        \textbf{Value} & 
        \textbf{Unit}\\
		\midrule
         R (cavity radius) & 120.00 & mm \\
         r (electrode radius) & 9.30 & mm \\
         d (electrode distance) & 18.60 & mm \\
         g1 (gap vert. vane $\leftrightarrow$ end plate) & 25.62 & mm \\
         g2 (gap horz. vane $\leftrightarrow$ end plate) & 8.35 & mm \\
         p (vane skirt position) & 60.0 & mm \\
         l1 (horizontal vane length) & 1353.07 & mm \\
         l2 (vertical vane length) & 1370.34 & mm \\
         L (cavity length) & 1378.69 & mm \\
         t (cavity thickness) & 20.0 & mm \\
         s (vane skirt max. thickness) & 30.0 & mm \\
         h (vane skirt min. thickness) & 10.0 & mm \\
        \bottomrule
    \end{tabular}
\end{table}

\subsubsection{RFQ re-buncher cell design}
\begin{figure}[b!]
	\centering
		\includegraphics[width=1.0\columnwidth]
        {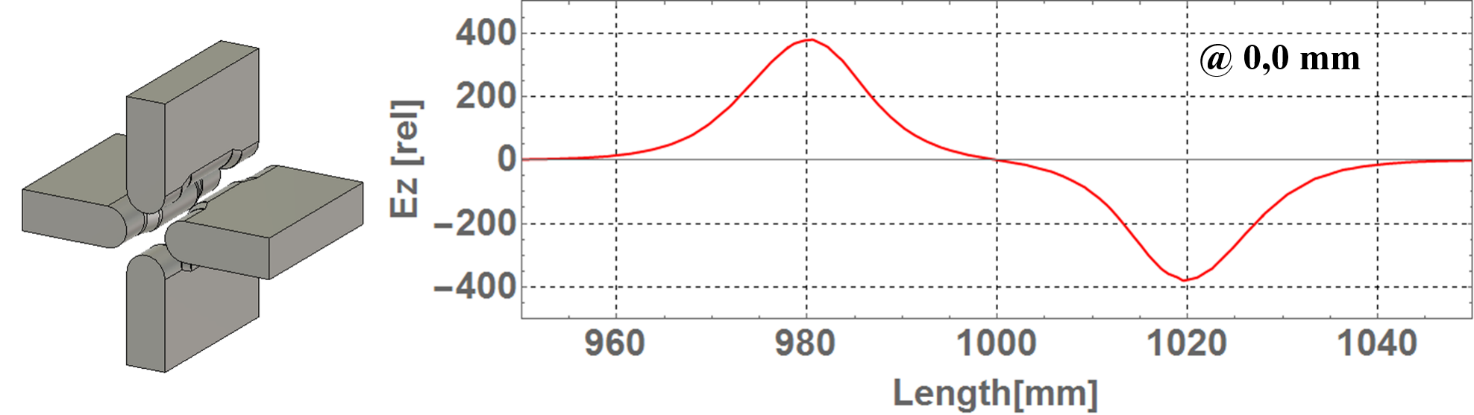}
	    \caption{3D CAD model of the trapezoidal re-buncher cell and 
                 associated field $E_z$ along the center axis.}
	    \label{fig:rfq_rebuncher}
\end{figure}
Finally, to achieve minimum energy spread and beam length at the matching point
(first accelerating cavity of the cyclotron), a re-buncher cell is added
to the RFQ, positioning the longitudinal beam focus about 40 cm after the RFQ exit.
Matching to the cyclotron will be discussed in more detail in \secref{sec:matching}.
The design parameters of the re-buncher are length, modulation, and position. 
The required length of re-buncher is calculated as \SI{78.94}{mm} based on the cell length, $\beta\cdot\lambda$. The modulation is determined to match the longitudinal 
focal length between RFQ and cyclotron. 
The position of the re-buncher in the RFQ is determined empirically to find
the optimum position with respect to phase and length of the bunch. 
The re-buncher uses a trapezoidal structure instead of a sinusoidal one 
to obtain an efficient field distribution and field separation from RFQ field,
as shown in \figref{fig:rfq_rebuncher}. 
The re-buncher is included in the CAD modeling of the vanes in Inventor.

%% file: Sec2.3_Cyclotron.tex
\subsection{Cyclotron Design \label{sec:cyclo}}
In the following subsections, we will briefly discuss the peculiarities of high intensity
cyclotrons and describe the IsoDAR compact cyclotron design. Most of the focus 
will be given to the design of the central region of the cyclotron and the matching
of the RFQ output beam to it. This will be done by discussing the 1 MeV/amu test cyclotron
that was designed for RFQ-DIP and which is a prototype of the central region of
the full IsoDAR cyclotron.

\subsubsection{Vortex Motion \label{sec:matching}}
In high current cyclotrons, the combination of space charge and the external 
focusing forces of the cyclotron's azimuthally varying main magnetic field can lead
to the formation of a stable, round (in the radial-longitudinal plane) bunch.
For the case of the PSI Injector 2 cyclotron, this has been extensively studied 
and discussed (see for example \cite{stammbach:vortex,yang:cyclotron_sim,baumgarten:vortex}).
The effect was dubbed \emph{vortex effect} or \emph{vortex motion} because the beam 
exhibits a vortex-like rotation in a local coordinate system (origin shifted to the bunch 
centroid position and local y-axis aligned with the bunch mean momentum).
It has been shown that, to achieve stable vortex motion, the bunches have to
be matched to the cyclotron through pre-bunching, focusing and carefully collimating 
the beam both before the cyclotron and inside, during the first few turns
\cite{yang:cyclotron_sim, stetson:strong_bunching, stammbach:strong_bunching}.
In this context, the following critical items are addressed for IsoDAR and 
RFQ-DIP through the measures described:
\begin{enumerate}
\item {\bf Longitudinal bunch size and phase acceptance:} The RFQ includes a trapezoidal re-bunching cell at the end (as described in the previous section)
to optimize the number of particles within the phase acceptance 
window of the cyclotron (typically $\pm 10\degree$ for isochronous machines)
during the crossing of the first accelerating cavity.
\item {\bf Transverse beam spread:} Firstly, the RFQ exit is brought very close 
to the entrance of the spiral inflector. Secondly, if necessary, there is room 
for a solenoid or quadrupole inside the cyclotron bore to provide additional
focusing. Third, the spiral inflector includes a special "V-shape" for 
additional focusing in the cyclotron axial direction (see \secref{sec:spiral}). 
\item {\bf Scraping of halo particles.} Particles that are outside of the phase acceptance of the cyclotron will inevitably become a nuisance, add to the beam
halo and reduce the turn separation at the extraction point. Careful placement 
of collimators to scrape these particles early on is done in highly
realistic simulations using the OPAL code \cite{adelmann:opal}.

\end{enumerate}
These parameters will be further optimized during the final design process.

\subsubsection{The full IsoDAR cyclotron}
\begin{figure}[t!]
\centering
	\includegraphics[height=0.5\columnwidth]{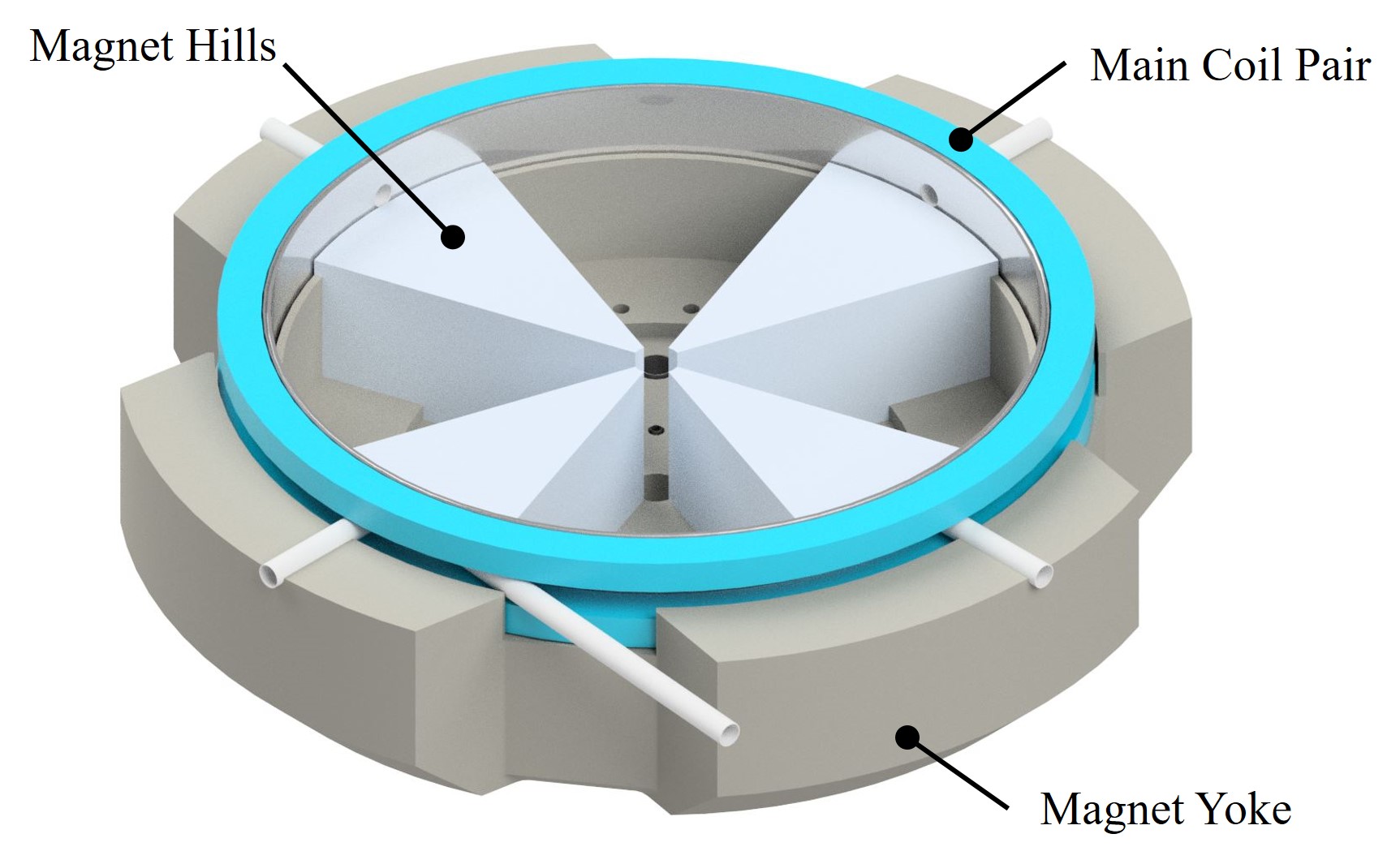}
	\caption{\footnotesize \label{fig:accel_mid_cut} 
             View of the bottom half of the full IsODAR cyclotron.
             The magnet yoke, magnet poles 
             (hill region) and cyclotron main coil pair are 
             indicated. Overall diameter: 6 m.}
\end{figure}

The design of IsoDAR cyclotron and beam dynamics simulations
starting during the third turn at 1.5 MeV/amu were presented in \cite{campo:isodar1, calanna:dic} and \cite{yang:daedalus}, respectively.
These reports were on the \DD Injector Cyclotron, which is identical to the
IsoDAR cyclotron except for the working frequency which is \SI{32.8}{MHz} (4\textsuperscript{th} harmonic) for IsoDAR and \SI{49.2}{MHz} (6\textsuperscript{th} harmonic) for the \DD injector cyclotron.
This cyclotron is a 4-sector machine with a pole radius of about 220 cm and 
is able to accelerate \htp ions from the injection energy of 35 keV/amu up to 
61.7 MeV/amu. The hill width varies from 25.5\degree at the inner radii up to
36.5\degree at outer radii. The hill gap is 100 mm, while the 
valley gap is very large (1800 mm) to be able to contain the RF cavities.
A half-section isometric view of the cyclotron can be seen in \figref{fig:accel_layout} and a mid-plane cut in \figref{fig:accel_mid_cut}.
The main studies up to now were focused on the space charge effects along the acceleration path and on the extraction of the beam using an electrostatic deflector \cite{yang:daedalus} using OPAL. In order to achieve sufficient turn separation 
at extraction, this cyclotron relies on the following:
\begin{enumerate}
\item {\bf High acceleration voltage.} A key feature of this machine is to 
use RF cavities able to produce accelerating voltages as high as 240 kV 
at the extraction region. This yields an inter-turn separation of about 
13 - 14.8 mm. 
\item {\bf Precessional resonance crossing in the extraction region.} 
A well-tuned fall-off of the magnetic field in the region of beam 
extraction leads to a crossing of the $\nu_r=1$ resonance (with $\nu_r$ the 
radial tune) a few centimeters before the extraction radius. This field configuration produces a precession of the accelerated orbit which increases the inter-turn separation to 20 mm.
\item {\bf Vortex effect.} This has been discussed in the previous section and 
collimator placements have been taken into account in \cite{yang:daedalus}. The vortex effect does not increase the inter-turn separation, but prevents
the beam from growing too large in the radial direction, which would lead to 
significant overlap even at high inter-turn separations.
\end{enumerate} 
According to the simulations performed with OPAL \cite{yang:daedalus}, taking into account the space charge forces in a beam bunch with phase length of 10\degree RF, the expected amount of beam loss on a 0.5 mm thick septum (the counter-part of the of the electrostatic deflector sitting between the last and second-to-last turn) should stay around 100 W for an extracted beam power of 600 kW. For safety reasons, we 
also plan to use four graphite stripper foils placed around the 
last orbit trajectory, to remove the remaining beam halo and to reduce the beam
power lost on the electrostatic deflector septum to virtually zero. 

\subsubsection{The RFQ-DIP 1 MeV/amu test cyclotron}
\begin{figure}[t!]
\centering
\includegraphics[height=0.5\columnwidth]{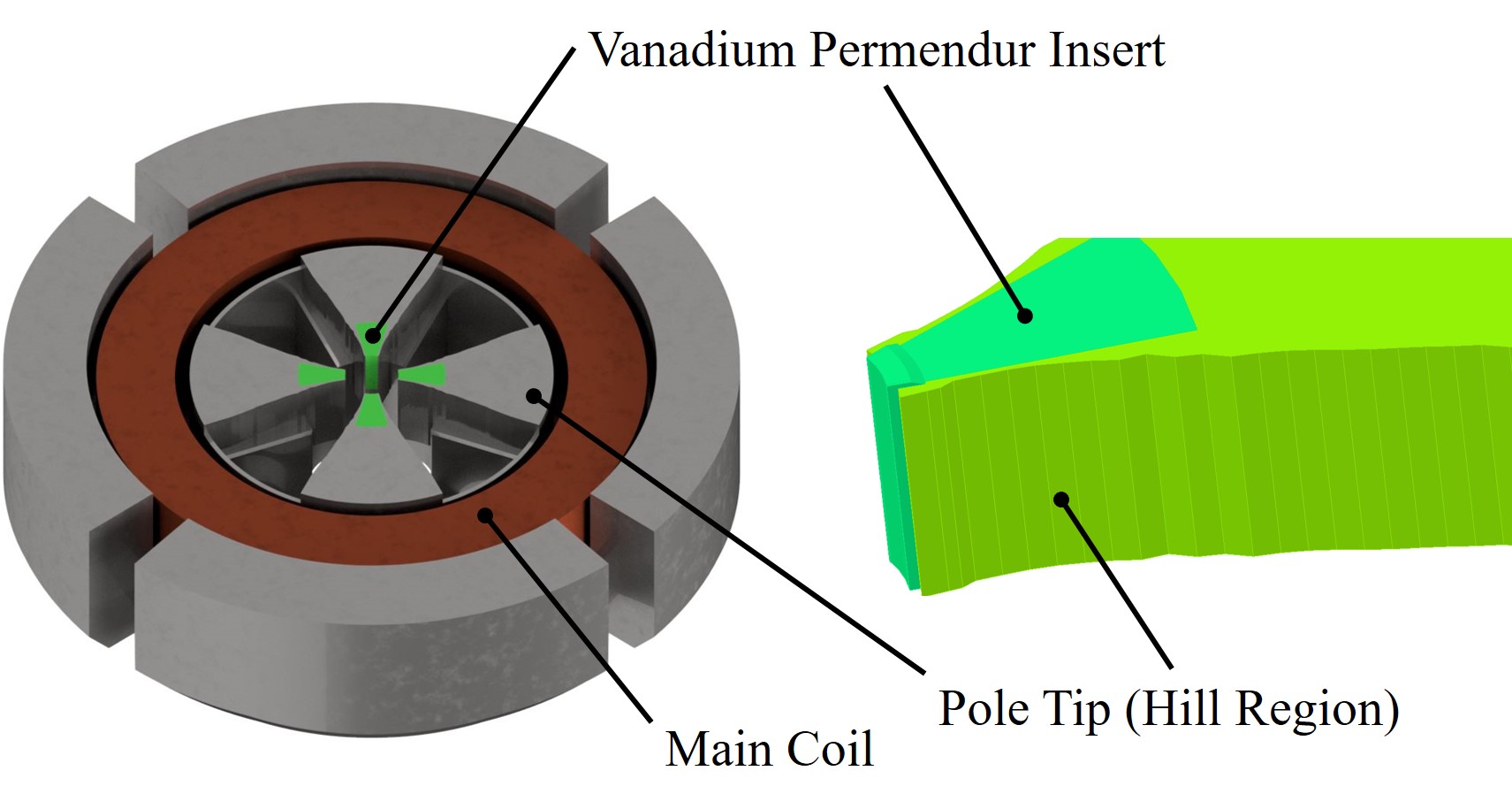}
\caption{\footnotesize \label{fig:perm2} 
View of the bottom half of the new 1 MeV/amu cyclotron design 
with the Vanadium Permendur insert (green). 
Overall diameter: 1.5 m}
\end{figure}

\begin{figure}[b!]
\centering
	\includegraphics[width=0.9\columnwidth]		
                    {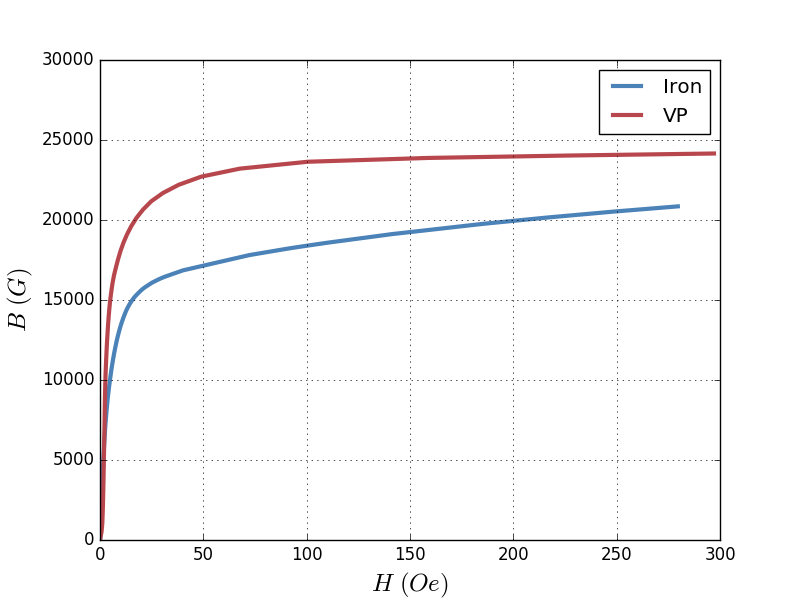}
	                \caption{\footnotesize Magnetic permeability
                    of soft iron and vanadium permendur (VP).
                    \label{fig:perm1}}
\end{figure}

As mentioned before, to test the central region and 
the acceptance of the beam therein, a separate \SI{1}{MeV/amu}
test cyclotron was designed with a magnetic field and clearances
for spiral inflector and collimators identical to the final
IsoDAR machine (see \figref{fig:perm2}).
Indeed, the previous design showed that the vertical focusing was not strong enough to focus the injected beam vertically. This 
was mainly due to the 10 cm vertical gap between the cyclotron 
magnet poles and due to the low flutter in the central region.
To increase the flutter (and hence the vertical focusing), the
pole gap was reduced and part of the iron of the hill in the central region was replaced with Vanadium Permendur (VP). Due to
the higher magnetic permeability of the VP (see \figref{fig:perm1}), it is possible 
to achieve higher magnetic fields in the hill region, which leads to a higher flutter of the field along the orbit trajectories of the particles and consequently better vertical focusing.

\begin{figure}[t!]
\centering
	\includegraphics[width=0.9\columnwidth] 
                    {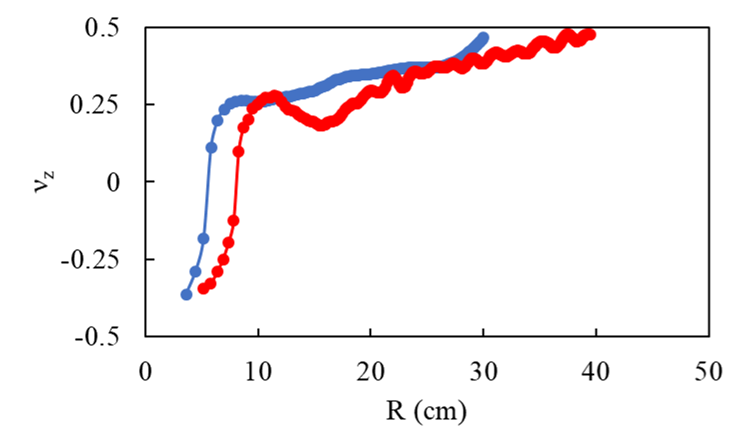}\\
	\includegraphics[width=0.9\columnwidth]
                    {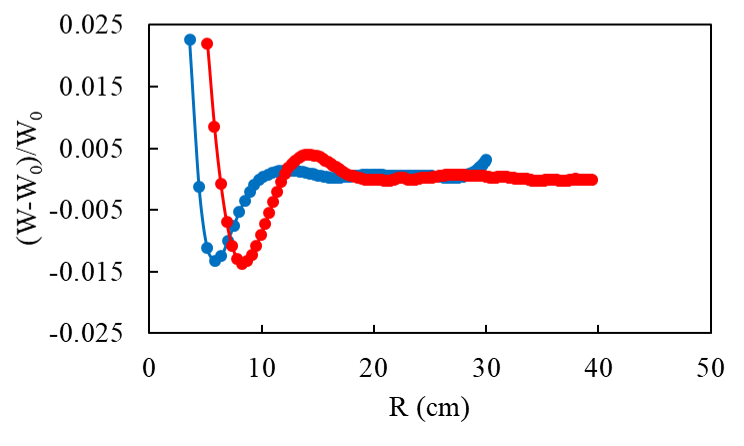}
	\caption{\footnotesize 
		     Upper: The $\nu_z$ values for the early model of 
             IsoDAR in red and for the new 1 MeV/amu test machine in blue. 
             Lower: Isochronism parameter $(W-W_0)/W_0$, for the early model 
             of IsoDAR in red and for the new 
             1 MeV/amu test machine in blue \label{fig:nuz_iso}}
\end{figure}

\begin{table}[!b]
    \centering
    \caption{1 MeV/amu test cyclotron design parameters. \label{tab:1mev_cyclo}}
    \begin{tabular}{lrl}
        \toprule
        \textbf{Parameter} & \textbf{Value} & \textbf{Unit}\\
		\midrule
            Yoke outer diameter & 1.5 & m \\
            Bore diameter & 27 & cm \\ 
            Weight & 6.3 & tons \\
            Pole angle & 23 to 46 & deg\\
            Pole radius & 37 & cm \\
            Coil cross-section & $16\times18$ & $\mathrm{cm}^2$ \\
            Max. field on the mid-plane & 1.48 & T \\
            Min. field on the mid-plane & 0.5 & T \\
			Energy of last closed orbit & 1.4 & MeV/amu \\
        \bottomrule
    \end{tabular}
\end{table}

The design parameters of this new test cyclotron are listed in 
\tabref{tab:1mev_cyclo}. The main items to highlight are the VP inserts
and the unusually large bore hole. The VP inserts start at a radius of 3.5 cm 
and end at 15 cm. They are protected by soft iron that can be removed for the fine shimming. A closer view is presented in \figref{fig:perm2}, right.
The bore hole has a diameter of 27 cm and allows insertion of the RFQ into the
cyclotron up to 20 cm from the mid-plane.

\begin{figure}[t!]
\centering
	\includegraphics[width=1.0\columnwidth]		
                    {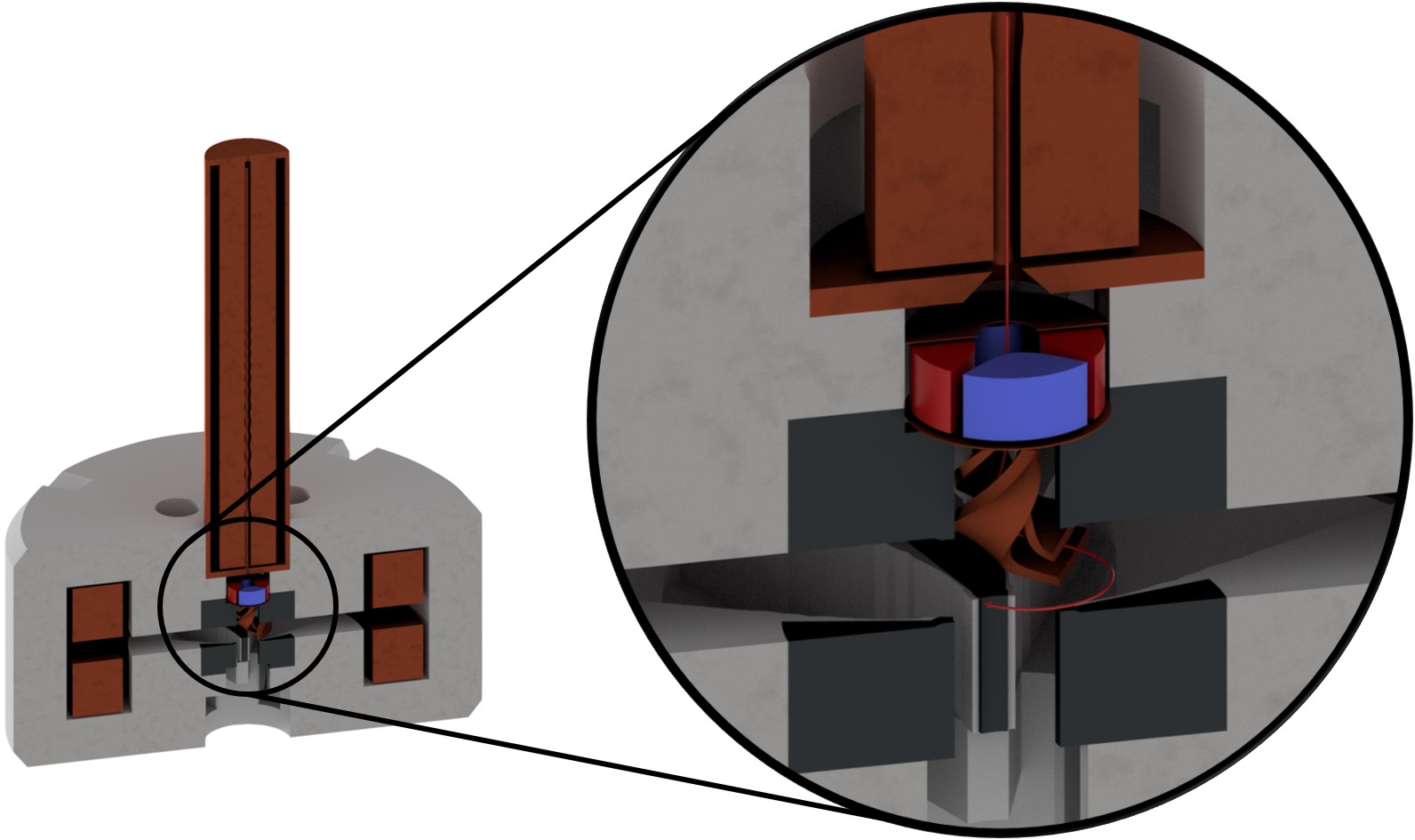}
	                \caption{\footnotesize CAD rendering of the central region of
                    the RFQ-DIP 1 MeV/amu test cyclotron. The follwing items have been
                    removed for visibility: Half the quadrupole entrance aperture, 
                    grounded housing of the spiral inflector, rf electrodes 
                    for acceleration in the cyclotron. An electrostatic quadrupole
                    singlet (electrodes shown in red and blue)
                    is placed between the RFQ and the spiral inflector 
                    for improved transmission and tunability.
                    \label{fig:1mev_full}}
\end{figure}

In \figref{fig:nuz_iso} (upper), the vertical tune $\nu_z$ is shown vs. radius for the IsoDAR cyclotron (red curve) and for the 1 MeV/amu test cyclotron. It is clearly visible
that the value of $\nu_z$ has increased  significantly at radii lower than 10 cm. 
This is important to achieve a sufficient vertical focusing along the first orbit where space charge effects are still strong.
In \figref{fig:nuz_iso} (lower), the isochronism parameter $(W-W_0)/W_0$
for the IsoDAR cyclotron and for the \SI{1}{MeV/amu} machine are shown. 
Both machines exhibit good isochronism and match very well in the range between 18 to 28 cm. This isochronism matching is important to guarantee that the beam accelerated in the \SI{1}{MeV/amu} machine could seamlessly be accelerated further in the 
\SI{60}{MeV/amu} IsoDAR machine.

As mentioned before, for the remainder of \secref{sec:accel_design}, all considerations
(transfer of the beam through the spiral inflector and RF electrode structure for the 
first few turns) will be with respect to the 1 MeV/amu RFQ-DIP test cyclotron.
The assembly of test cyclotron, RFQ and the matching elements (electrostatic quadrupole
singlet and spiral inflector) are shown in \figref{fig:1mev_full}.
A single electrostatic quadrupole has been found to give a measure of 
tunability and increases the transmission of the beam through the spiral inflector
from roughly 70\% to 80\%.

%% file: Sec2.4_Spiral.tex
\subsubsection{Design considerations for spiral inflectors \label{sec:spiral_code}}
Many of the spiral inflector design parameters depend on the beam and particle properties. Primarily, there are two free parameters: the applied electric field, and a tilt parameter, which is denoted by $k^{\prime}$. Electrode voltages and electrode gap width determine the electric bending radius of the spiral inflector, which corresponds exactly to the height of the device. $k^{\prime}$ describes an angle at which the exit of the electrodes is tilted. This parameter can be changed to align center of curvature for the particle trajectories. In the analytical treatment of the spiral inflector, the magnetic field in the central region is taken to be constant in the axial direction. This assumption allows for an exact solution of a central ion trajectory through the spiral inflector which can be used to calculate the surfaces of the spiral electrodes \cite{toprek:spiral}.

For taller spiral inflectors, it may not be appropriate to assume the magnetic field is constant. The magnetic field of a cyclotron along the z-axis may vary significantly enough such that the beam does not follow the analytical central trajectory. Secondly, the fringing electric fields at the entrance and exit of the spiral inflector cause the beam to exit with an angle to the mid-plane of the cyclotron. To account for both the fringing electric field and a non-uniform magnetic field, numerical analyses and optimization routines have been developed in a {\verb python } module to aid in the design process of spiral inflectors \cite{weigel:spiral}. This code applies the analytical treatment of the spiral inflector and performs a full electric field calculation, which includes the fringe fields using a boundary elements method implemented in the {\verb python } module BEMPP \cite{smigaj:bempp}. A design particle is then tracked through the spiral inflector and
the entrance and exit of the spiral inflector are modified to compensate for the effect
of fringe fields.

\subsubsection{The IsoDAR/RFQ-DIP spiral inflector \label{sec:spiral}}

\begin{figure}[t!]
\centering
\centerline{\includegraphics[width=0.7\columnwidth]{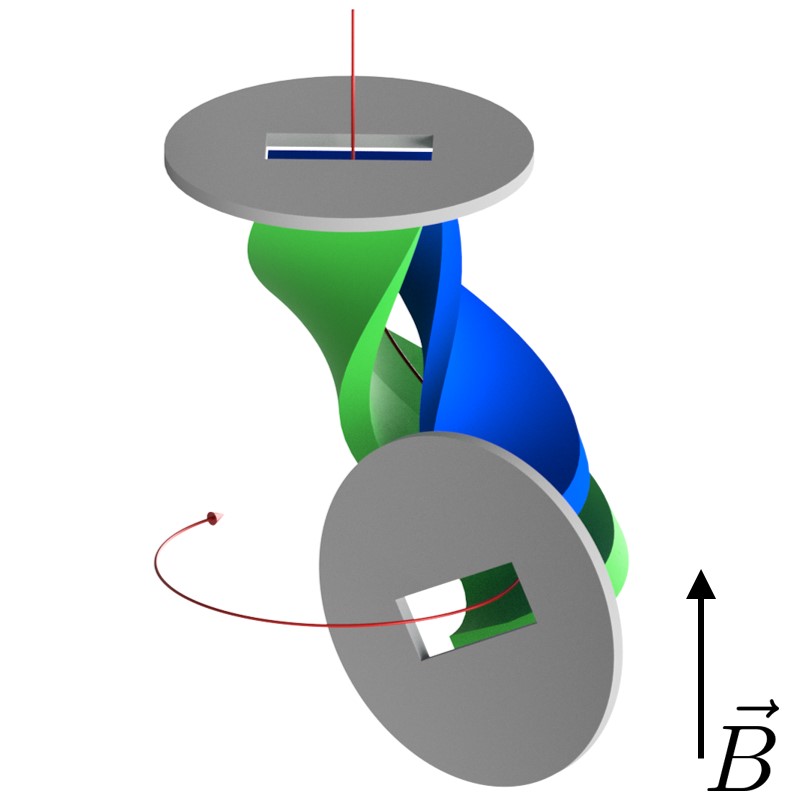}}
\caption{3D view of the IsoDAR spiral inflector (anode: green, cathode: blue) and apertures (gray) with a reference \htp trajectory (red).}
\label{IsoDAR_inflector_Weigel}
\end{figure}

\begin{table}[!b]
    \centering
    \caption{Spiral inflector design parameters. \label{tab:spiral_inflector}}
    \begin{tabular}{lrl}
        \toprule
        \textbf{Parameter} & \textbf{Value} & \textbf{Unit}\\
		\midrule
            Electrode voltages & $\pm$12 & kV \\
		    Input energy & 70 & keV \\
            Electrode width & 1.0 & cm \\
            Gap distance & 1.8 & cm \\
            Aspect ratio & 2.5 & \\
            Tilt angle & 27 & deg \\
        \bottomrule
    \end{tabular}
\end{table}

A preliminary design of the IsoDAR spiral inflector, including entrance and exit apertures, is presented in Figure \ref{IsoDAR_inflector_Weigel}.
The geometry of the spiral inflector was generated using the newly developed code presented in Section \ref{sec:spiral_code}, using a numerical analysis to correct 
for fringe field effects and a non-uniform magnetic field. 
The aspect ratio between the cross-sectional width and length of the spiral electrodes was set to be 2.5.
The opening of the grounded apertures were designed to match the shape of the entrance and exit of the spiral inflector to shield the incoming particles from the strong fringing electric fields and to collimate the beam.
The top aperture is placed 5 mm from the entrance of the spiral inflector and the bottom aperture is placed 10 mm from the exit, each with a thickness of 
\SI{4}{mm}.

The high beam current that is being delivered to the cyclotron necessitates a large initial gap distance of \SI{1.8}{cm} to minimize the loss of particles during injection. Additionally, the voltage assigned to the electrodes was chosen to be $\pm$12 kV. The IsoDAR spiral inflector is taller than a typical spiral inflector because of the gap distance since the height is inversely proportional to the strength of the electric field between the electrodes.

\subsubsection{The IsoDAR Central Region}
\begin{figure}[!t]
\centering
\centerline{\includegraphics[width=200pt]{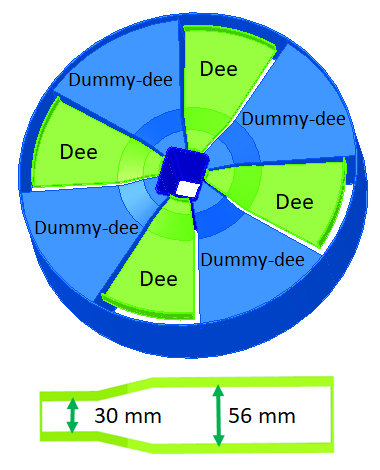}}      
\caption{3D view of the IsoDAR central region. The dummy-dees are in blue while the dees are in green.}
\label{IsoDAR_CR_Grazia}
\end{figure}
The final element of the RFQ-DIP test cyclotron is the RF system consisting of
dees and dummy-dees. A rendering of the preliminary design is shown in
\figref{IsoDAR_CR_Grazia}.
After leaving the spiral inflector, the \htp ions are injected into the cyclotron central region at 70 keV energy and accelerated up to the 1 MeV/amu. The accelerating structure is composed of four electrodes, the so-called \emph{dees}, driven by a RF voltage with a maximum of 70 kV and by as many electrodes at ground, usually named \emph{dummy-dees}.

The RF resonance frequency of the cavities is 32.8 MHz and the accelerating harmonic is the $\mathrm{4^{th}}$, four times the revolution frequency of the beam in the magnetic field of the IsoDAR cyclotron. The vertical aperture of the electrodes has two different values, 30~mm in the first turn and 56~mm after. The choice of smallest vertical gap at the inner cyclotron radii allows to reduce the transit time factor effects on the beam dynamics in the first turn. \\
Finally, the angular width of the dees is 36 degrees and the gap between each dee and dummy-dee is 12 mm.

While the preliminary results of the RF system are promising (as reported in 
\secref{ssec:spiral_sim}), further optimization is necessary, 
and a detailed design study performed by the AIMA company in France
is currently on the way.

%% file: Sec3_Simulations.tex
\begin{figure}[t!]
	\centering
		\includegraphics[width=0.9\columnwidth]
        {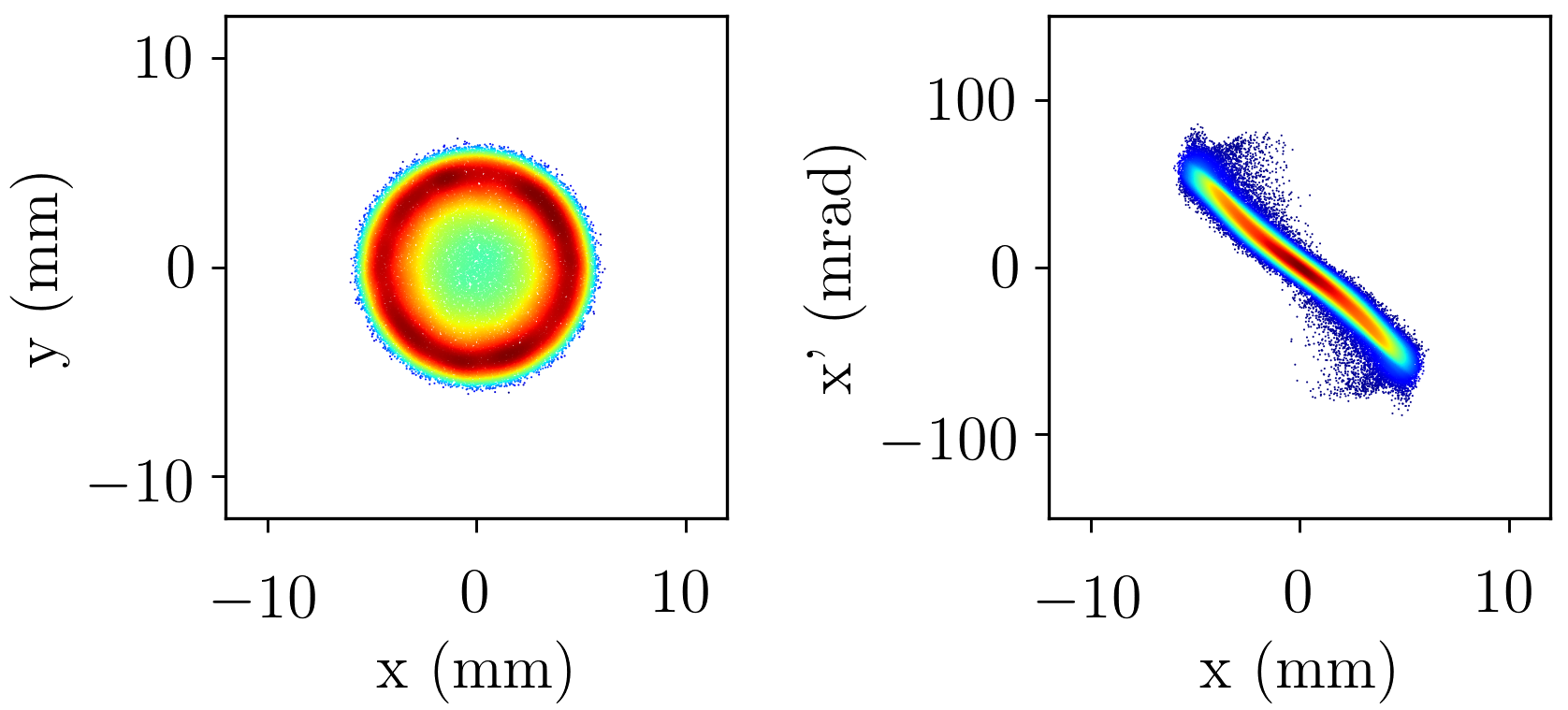}
	    \caption{Beam cross-section (left) and xx' phase space (right)
                 for 8 mA of \htp (transported together with 2 mA of protons)
                 at the entrance flange of the RFQ. The beam is converging so
                 that the transverse focus lies well within the RFQ's 
                 \emph{radial matching section}. Corresponding beam parameters
                 are listed in \tabref{tab:init_distr}.}
	\label{fig:init_distr}
\end{figure}
\section{Simulations\label{sec:simulations}}
In the following section, the latest simulation results of the RFQ-DIP design
will be described briefly. Simulations showing the feasibility of the
IsoDAR main cyclotron design by accelerating high intensity beams from 
later turns have been reported in \cite{yang:daedalus} 
(starting at \SI{1.5}{MeV/amu}) and \cite{jonnerby:thesis} (starting at 
\SI{193}{keV/amu}) and are not subject of this manuscript. A detailed report 
on the full start-to-end simulations of RFQ-DIP and IsoDAR is forthcoming.

\subsection{Initial conditions from ion source}
The initial conditions for the RFQ simulations are obtained from 
simulations of the ion source plasma extraction using IBSIMU, as 
was described in \secref{sec:source}. 
We consider two cases: the nominal case of 10 mA total extracted beam current
and the exaggerated case of 20 mA total extracted beam current 
(both with 80\% \htp, 20\% protons).
The beam parameters at the entrance flange of the RFQ (rightmost electrode in
\figref{fig:extraction}) are listed in \tabref{tab:init_distr} and a
representative phase space is shown in \figref{fig:init_distr}.

\begin{table}[!b]
    \centering
    \caption{Initial beam parameters for RFQ simulations obtained                  
             from IBSIMU calculations. (only the horizontal direction 
             is shown, the vertical direction is similar).
             \label{tab:init_distr}}
    \begin{tabular}{lcccc}
        \toprule
        &  \multicolumn{2}{c}{10 mA} & \multicolumn{2}{c}{20 mA}\\
        & protons & \htp & protons & \htp \\
		\midrule
        	Fraction (\%) & 20 & 80 & 20 & 80 \\
            Current (mA) & 2 & 8 & 4 & 16\\
        	Energy (keV) & 15 & 15 & 15 & 15 \\
        	$\diameter_\mathrm{2rms}$ (mm) & 11.8 & 11.8 
                                           & 14.4 & 14.4 \\
            $\epsilon_{\mathrm{x, n, 4rms}}$ ($\pi$-mm-mrad) & 0.60 & 0.43 
                                                             & 0.91 & 0.65 \\
            4rms includes (\%) & 90.7 & 90.7 
                               & 88.4 & 88.3\\
        \bottomrule
    \end{tabular}
\end{table}

\subsection{RFQ simulations}
\begin{figure}[t!]
	\centering
		\includegraphics[width=0.9\columnwidth]
        {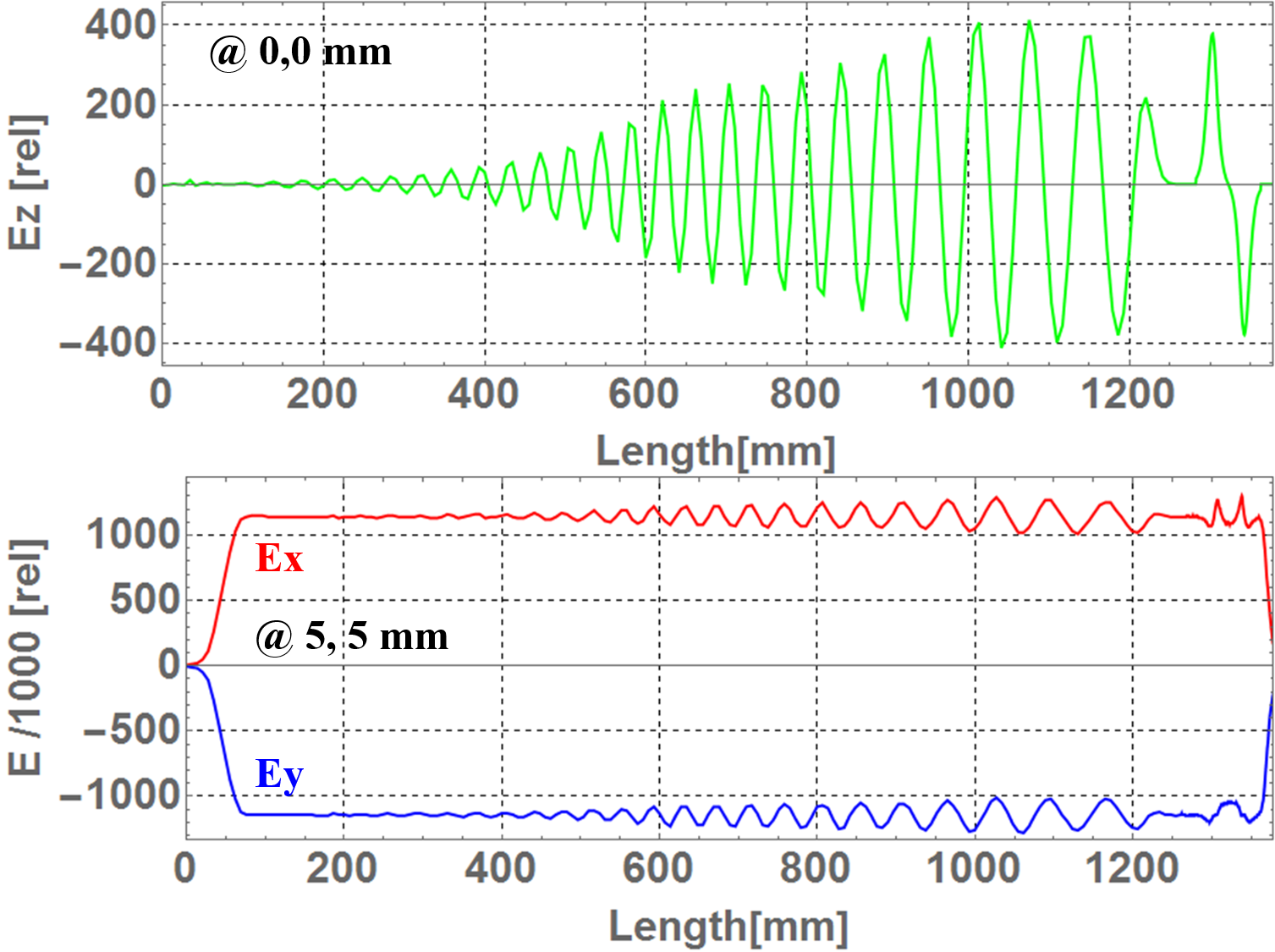}
	    \caption{The field distribution along beam axis with re-buncher.}
	\label{RFQ_rebuncherfield}
\end{figure}

As a part of the RFQ-DIP and IsoDAR systems, the RFQ accelerator provides 
acceleration of the ion beam generated from the ion source and highly efficient 
bunching as well as beam separation by different mass-to-charge ratios 
(1 for the proton beam and 2 for the \htp ion beam).
In this section, we described the beam dynamics in the RFQ accelerator using the particle distributions from the previously reported IBSIMU results. 
The electric field distributions along the beam axis and 5 mm offset are presented in \figref{RFQ_rebuncherfield}. These include the re-buncher fields in the transition cell of the designed IsoDAR RFQ, to obtain additional longitudinal focusing. 

\begin{figure}[t!]
	\centering
		\includegraphics[width=1.0\columnwidth]
        {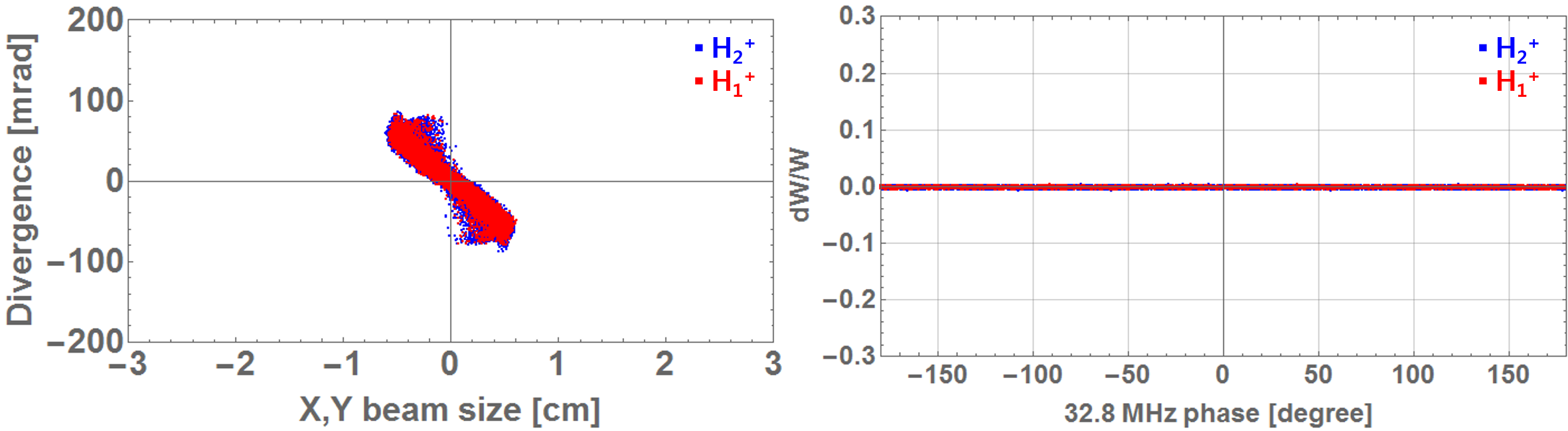}
        \includegraphics[width=1.0\columnwidth]
        {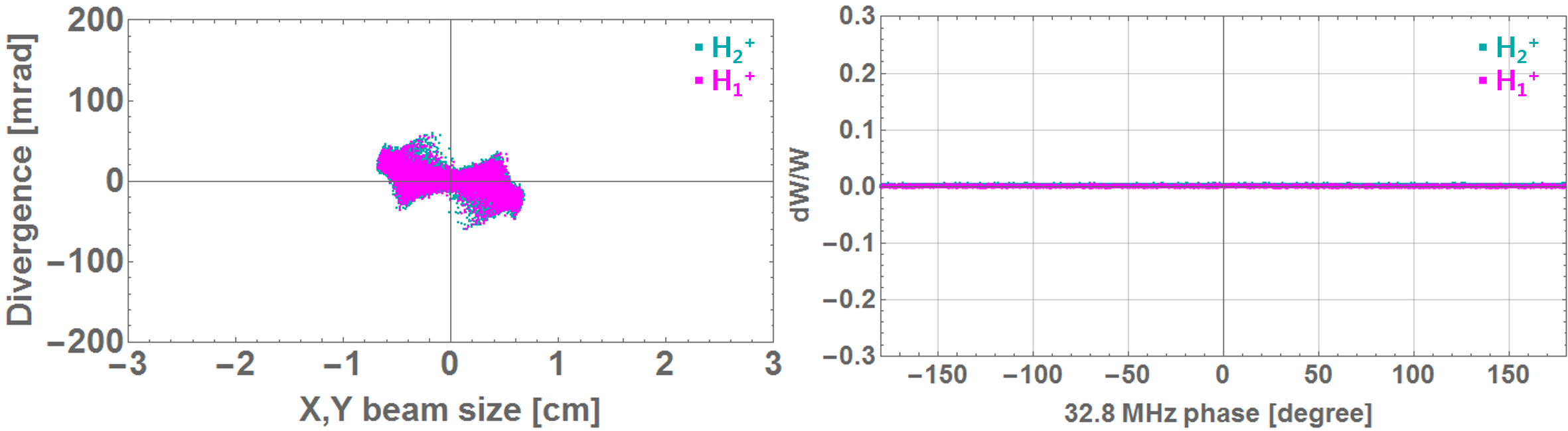}
         \caption{The transverse and longitudinal beam distributions at the entrance of the IsoDAR RFQ in cases of 10 mA (top) and 20 mA (bottom).}
	\label{ste_rfq_initial}
\end{figure}

\begin{figure}[t!]
	\centering
		\includegraphics[width=1.0\columnwidth]
        {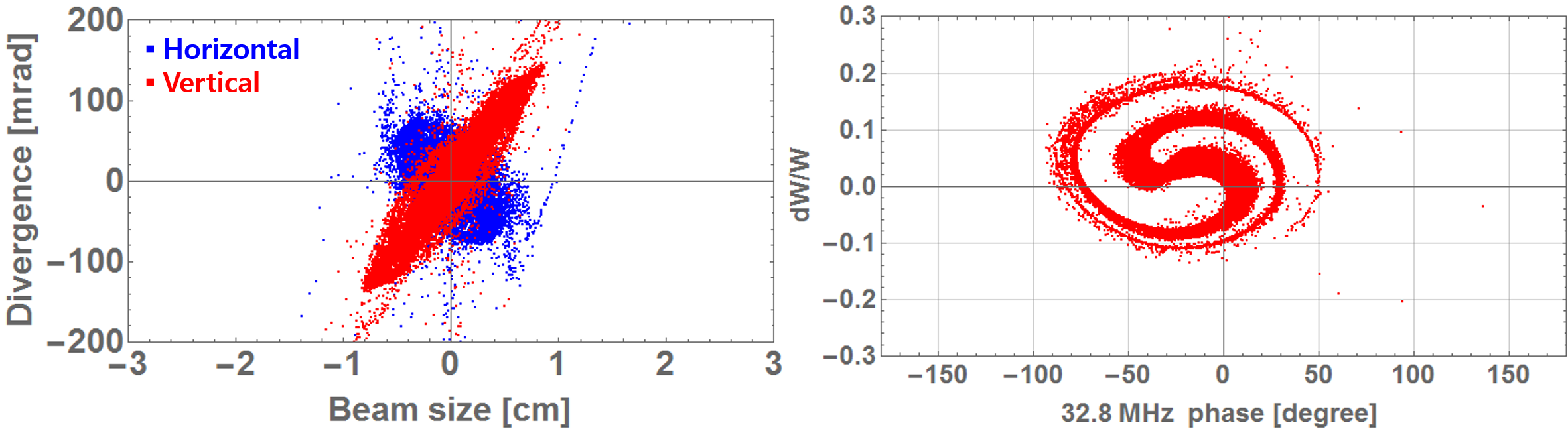}
        \includegraphics[width=1.0\columnwidth]
        {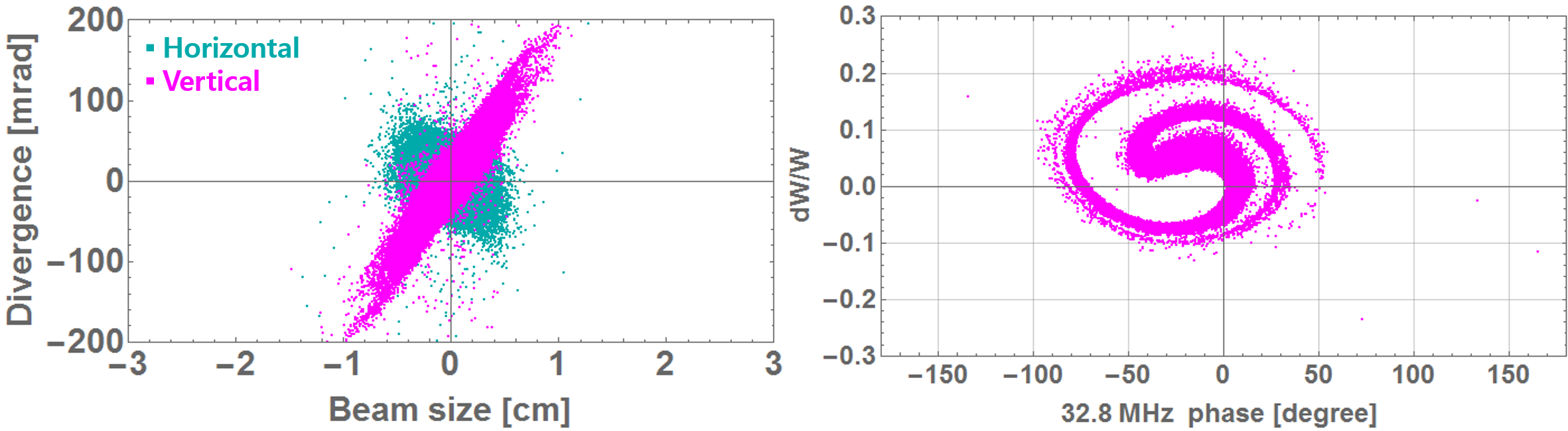}
         \caption{The transverse and longitudinal beam distributions at the 
         exit of the IsoDAR RFQ for the two cases of 10 mA (top) and 20 mA 
         (bottom).}
	\label{ste_rfq_outputbds}
\end{figure}

\begin{figure}[t!]
	\centering
		\includegraphics[width=1.0\columnwidth]
        {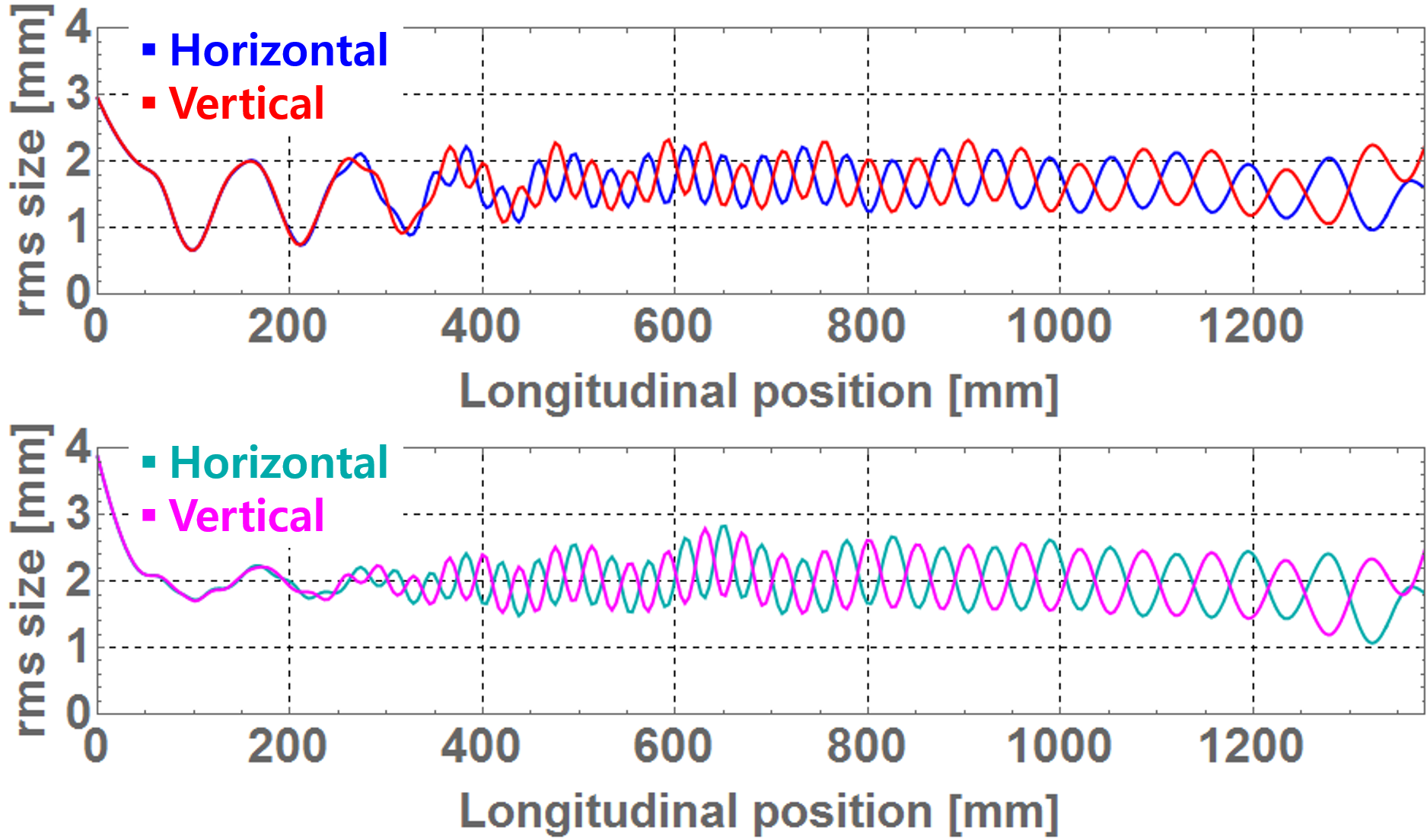}
         \caption{The RMS beam size along the RFQ cavity for the two cases 
                  of 10 mA (top) and 20 mA (bottom).}
	\label{ste_rfq_envelope}
\end{figure}

As an operational scenario, studies with 10 mA and 20 mA of beam intensity have been performed. The initial beam distributions of the two cases at the entrance of the IsoDAR RFQ are presented in \figref{ste_rfq_initial}. Note that the upper row in 
\figref{ste_rfq_initial} corresponds directly to the distribution shown in
\figref{fig:init_distr}.
The generated particles from the ion source, including additional, unwanted species,
such as protons and \hthp, need to be separated and selected by the RFQ.
When particles with the wrong mass-to-charge ratio are injected into the RFQ, 
they will not get accelerated, due to the improper RF frequency fields. 
Using the injected beam distributions from the ion source, the results of the RFQ simulations for both cases have transmission efficiencies above 97\% and 8.186 and 6.908 keV/u-ns of longitudinal emittance, respectively. 3D tracking was 
done using the TRACK code.
The RMS envelopes of transverse beam size for the two cases of 10 mA and 20 mA of beam intensity are presented in \figref{ste_rfq_envelope}. They are smaller than 2.96 mm for 10 mA and 3.88 mm for 20 mA, respectively. The beam distributions at the end of the RFQ are shown in \figref{ste_rfq_outputbds}. As expected, the transverse emittances increase with beam intensity, due to space charge effects. 
The detailed simulation results for the two cases are summarized in \tabref{ste_rfq_results}.

\begin{table}[!b]
    \centering
    \caption{Important resulting parameters of the RFQ simulations for the 
             current RFQ-DIP/IsoDAR design (r=rms, n=normalized).
             \label{ste_rfq_results}}
    \begin{tabular}{llll}
        \toprule
        Elements        & 10 mA & 20 mA & units  \\
		\midrule
Transmission rate	& 98.48	& 97.12	& \% \\
Input tr. emit (r,n)	&0.106	&0.304	&mm-mrad  \\
Output emit x (r,n)	&0.306	&0.373	&mm-mrad  \\
Output emit y (r,n)	&0.329	&0.393	&mm-mrad  \\
Output emit z (r)	&8.186	&6.908	&keV/u-ns  \\
        \bottomrule
    \end{tabular}
\end{table}

\subsection{Spiral inflector and central region simulations \label{ssec:spiral_sim}} 
The simulation studies of the IsoDAR spiral inflector and central region
reported here, were performed in two steps:
\begin{enumerate}
\item Preliminary simulation study in collaboration with the company IBA in Belgium.
\item Detailed 3D simulation study including space charge (ongoing effort at MIT and
the company AIMA in France).
\end{enumerate}

\paragraph{Preliminary simulations}
The results of the preliminary simulations can be found here \cite{CR} and are 
briefly summarized below.
By using an iterative approach, the spiral inflector and central region have been designed and simulated to guarantee good injection and acceleration  of the reference orbit in the test cyclotron center.
The motion of a single particle, starting from a position along the cyclotron vertical axis (100 mm from the median plane, 20 mm from the spiral inflector
entrance) was tracked. The initial RF phase of the reference orbit was selected with the aim to optimize the vertical excursion in the cyclotron center, the horizontal orbit centering and the energy gain per turn.
The designed spiral inflector and central region provide a good injection and acceleration of the reference orbit in the test cyclotron. 
To test transmission, 5000 particles with the same distribution in the both x-x$^\prime$ and y-y$^\prime$ phase planes were then simulated. A normalized emittance of 1 $\pi\cdot$mm$\cdot$mrad was used, a value that is about 20\% larger than the expected beam emittance delivered by the RFQ. A round, uniform beam 
with \SI{10}{mm} diameter has been considered.
In addition an energy spread equal to $\pm 2\%$ has been taken into account. The results of the simulations show that losses are localized in the cyclotron center, 
in particular at the inflector exit and in the first half turn, since the vertical aperture of the housing hole and electrodes is smaller than beam vertical envelope. This is due to the growth of beam emittance through the spiral inflector and the transverse defocusing of the beam at the inflector exit. The space charge effects play an important role in the beam growth and transverse defocusing determining an increase of the losses. In addition, turn separation has been 
observed to be too small, due to the overlap between the tails of the beam distribution of adjacent turns. 
Further optimization and inclusion of the actual RFQ output beam distribution was
deemed necessary.

\begin{figure}[t!]
	\centering
		\includegraphics[width=0.9\columnwidth]
        {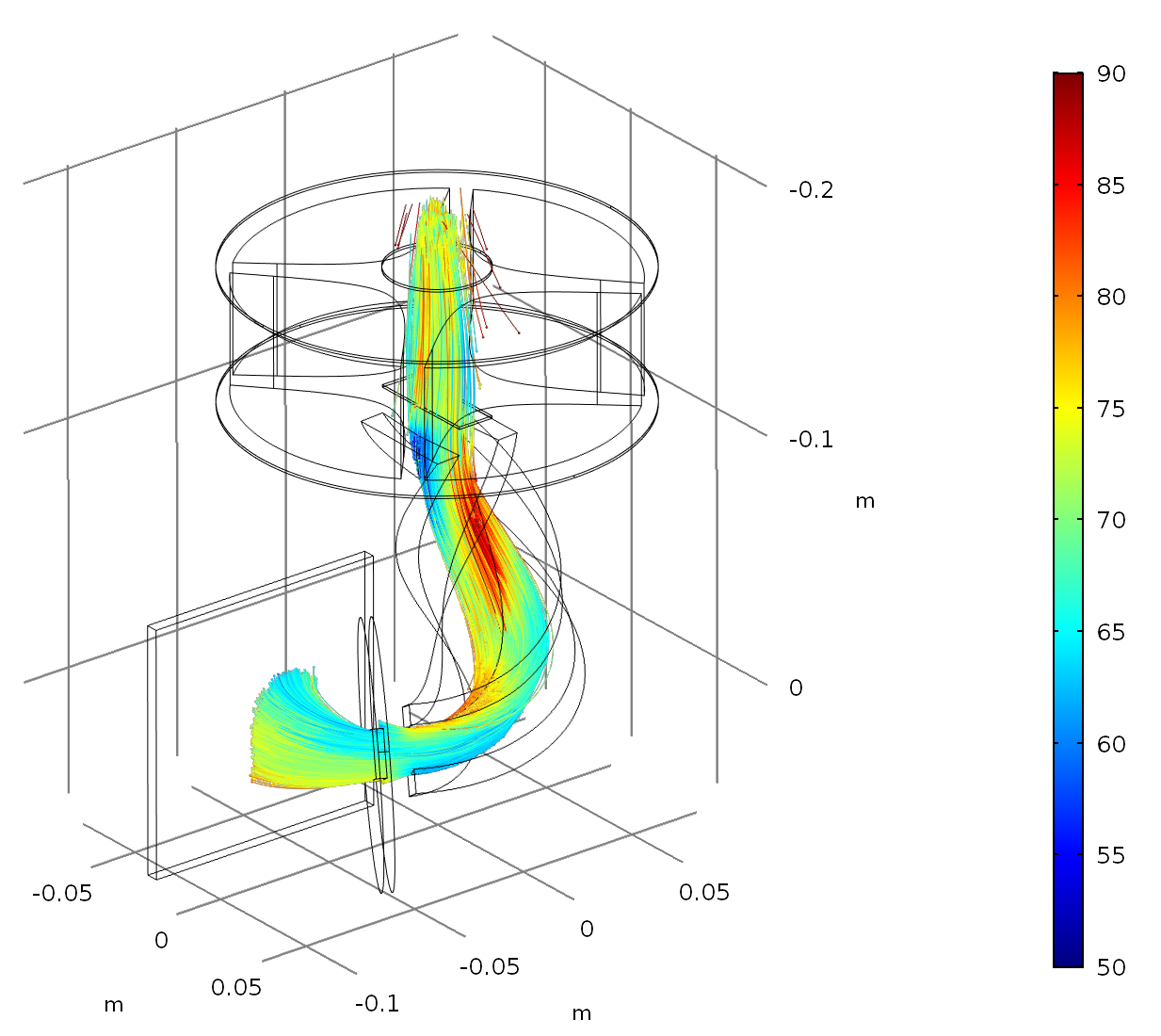}
        \caption{The RFQ output particle distribution tracked through the 
                 electrostatic quadrupole singlet and the spiral
                 inflector, using COMSOL Multiphysics. the particles are 
                 terminated on a beam stop immediately after exiting the inflector.
                 Colorbar: particle kinetic energy (keV).}
	\label{fig:si_comsol}
\end{figure}

\paragraph{Detailed simulations}
the first step in the detailed simulation study was to generate a spiral inflector
using the methods described in \secref{sec:spiral_code}. The electrostatic quadrupole
singlet was added and the fields were calculated in COMSOL Multiphysics. The RFQ
output particles presented in the previous subsection were then tracked, including space charge, for 10 mA and 20 mA of beam. The result of the 20 mA case is shown
in \figref{fig:si_comsol}. The beam was well-centered on the mid-plane with 
\SI{12.3}{kV} applied to the spiral electrodes and has a maximum vertical extent 
of \SI{35}{mm}. The transmission in both cases was around 80\% when using 
\SI{11.5}{kV} for the quadrupole. Most of the losses occur towards the end of 
the inflector due to the initial energy spread of the beam. Further reduction
of the beam energy spread can be achieved with a vertical aperture close to the 
exit of the inflector, albeit at the expense of higher losses.
These particles are now being used in the optimization of the central region
by AIMA. This process includes shimming the magnetic field to reduce first harmonic 
perturbations, shaping the dees and placing pillars (collimators) along the beam path in the central region. This removes the beam halo radially in order to achieve well separated turns. The use of pillars will also allow to increase the electrode vertical aperture to reduce the vertical losses and the transit time factor will be mitigated by the smaller radial aperture for the beam.

As mentioned above, this simulation effort is ongoing and another paper on the 
full start-to-end simulations of the RFQ-DIP and IsoDAR projects is forthcoming.

%% file: Sec4_Conclusion.tex
\section{Summary and Outlook \label{sec:outlook}}
The prospect of being able to reliably create and deliver cw
proton beam currents of 10 mA and more at energies upwards of 60 MeV
is very exciting for the fields of particle physics (e.g. neutrino 
experiments \cite{IsoDARsterilePRL, abs:isodar_cdr1, abs:daedalus, aberle:daedalus}), medical isotope production \cite{schmor:isotopes,alonso:isotope}, 
and materials research (see for example \cite{lester:nrl_report}). 
A system that can perform at those levels was described in this paper. 
It consists of a high intensity \htp ion source, RFQ injection system, and compact
cyclotron for acceleration. 
As was discussed here, the main challenges are controlling space charge and achieving high injection efficiency into the cyclotron. We have shown preliminary designs and simulations showing that both are possible using 
\htp as the primary ion, an RFQ to aggressively pre-bunch the beam before injecting it through a spiral inflector, and by carefully designing a central region that facilitates 
vortex motion and uses collimators to clean up beam halo.
The beam was matched to the central region of
the cyclotron and acceleration for 4 turns showed good turn separation as well as the onset of vortex motion, which has a stabilizing effect on the beam. 
The next steps on the path to a widely usable high intensity proton driver, will
be the successful completion of the RFQ-DIP, demonstrating RFQ direct injection
into a 1 MeV/amu test cyclotron,
followed by the full IsoDAR machine, which will be able to deliver 600 kW of beam
power (10 mA cw protons at 60 MeV after a stripper foil). 
As was described in \secref{sec:accel_design}, the RFQ - Direct Injection Project is well underway and we expect completion of the 
test bench and first results by the end of 2019.
A proposal for the full IsoDAR system will be submitted by the end of 2018.

Once the high intensity \htp beam has been extracted from the 
compact cyclotron, one can envision to inject it into another cyclotron to boost the beam energy even higher. This process is described in detail elsewhere \cite{abs:daedalus, aberle:daedalus} for the case of
 (\DD), another neutrino experiment searching for cp-violation. 
Here a separated sector superconducting cyclotron is used to accelerate \htp from 60 MeV/amu to 800 MeV/amu. However, one does not have to stop at 800 MeV/amu, and with the technology developed 
here, multi-megawatt compact systems using two chained cyclotrons
are now within reach for energy research (needs described for example in \cite{ishi:adsr,rubbia:adsr,biarrotte:ads,lisowski:ads}).